\documentclass{scrartcl}
 
\RequirePackage{fix-cm}
\RequirePackage[utf8]{inputenc}
\RequirePackage[T1]{fontenc}
\RequirePackage{xspace}
\RequirePackage{amssymb}
\RequirePackage{amsmath}
\RequirePackage{amsthm}
\RequirePackage[hidelinks]{hyperref} 
\RequirePackage[round]{natbib}

\RequirePackage{nicefrac}
\RequirePackage{tikz}
\usetikzlibrary{positioning,arrows,shapes,decorations,calc,fit,arrows.meta} 
\RequirePackage{colortbl}
\RequirePackage{booktabs}
\RequirePackage{cleveref}
\RequirePackage{array}
\RequirePackage{graphicx}
\RequirePackage{xcolor}
\RequirePackage{enumitem}
\RequirePackage{tabularx}

\theoremstyle{definition}
\newtheorem{definition}{Definition}

\theoremstyle{plain}
\newtheorem{theorem}{Theorem}
\newtheorem{lemma}{Lemma}
\newtheorem{proposition}{Proposition}
\newtheorem{corollary}{Corollary}

\newtheorem{observation}{Observation}

\newtheoremstyle{bfnote}
{}{}%
{}{}%
{\bfseries}{.}%
{ }%
{\thmname{#1}\thmnumber{ #2}\thmnote{\textnormal{ (#3)}}}
\theoremstyle{bfnote}
\newtheorem{remark}{Remark}

\newcommand{\Omit}[1]{} 
\usepackage{thm-restate}
\usepackage{subcaption}
\usepackage{tikz-cd}
\usepackage{mathtools}
\DeclareMathOperator*{\argmax}{\arg \max}
\newcommand{\supp}{\mathrm{supp}} 

\newcommand*{\mask}[2]{%
    \mathord{\makebox[\widthof{\(#1\)}]{\(#2\)}}%
}

\newcommand{\an}[1][]{\ifthenelse{\equal{#1}{}}{anonymity\xspace}{anonymity\xspace}}
\newcommand{\neu}[1][]{\ifthenelse{\equal{#1}{}}{neutrality\xspace}{neutrality\xspace}}
\newcommand{\strp}[1][]{\ifthenelse{\equal{#1}{}}{strategyproofness\xspace}{strategyproofness\xspace}}
\newcommand{\prop}[1][]{\ifthenelse{\equal{#1}{}}{proportionality\xspace}{proportionality\xspace}}
\newcommand{\rr}[1][]{\ifthenelse{\equal{#1}{}}{range-respecting\xspace}{range-respecting\xspace}}
\newcommand{\eff}[1][]{\ifthenelse{\equal{#1}{}}{efficiency\xspace}{efficiency\xspace}}

\newcommand{\edr}[1][]{\ifthenelse{\equal{#1}{}}{EDR\xspace}{EDR(#1)\xspace}}

\newcommand{\nash}[1][]{\ifthenelse{\equal{#1}{}}{\ensuremath{\mathit{NASH}}\xspace}{\ensuremath{\mathit{NASH}(#1)}\xspace}}

\newcommand{\p}{\textbf{p}} 
\newcommand{\q}{\textbf{q}} 
\newcommand{\s}{\textbf{s}} 
\newcommand{\med}{\mathrm{med}} 

\newcommand{\profile}[7]{\captionlistentry{}
\centering

\begin{tabular}{m{0.03\textwidth} c c c}
&\multicolumn{3}{c}{\footnotesize Profile #1}\\
& a & b & c \\
& #2 \\
& #3 \\
& #4 \\
\cmidrule{2-4}
$\q^{(#1)}$ &$\mask{4/15}{#5}$&$\mask{4/15}{#6}$&$\mask{4/15}{#7}$
\end{tabular}
}

\newcommand{\profileagent}[7]{\captionlistentry{}
\centering

\begin{tabular}{r c c c}
&\multicolumn{3}{c}{\footnotesize Profile #1}\\
\# agents& a & b & c \\
1& #2 \\
$n-2$& #3 \\
1& #4 \\
\cmidrule{2-4}
$\q^{(#1)}$ &$\mask{\ge 1/2n}{#5}$&$\mask{ \ge (2n-5)/2n}{#6}$&$\mask{4/15}{#7}$
\end{tabular}
}

\newcommand{\profileagentmod}[7]{\captionlistentry{}
\centering

\begin{tabular}{r c c c}
&\multicolumn{3}{c}{\footnotesize Profile #1}\\
\# agents& a & b & c \\
1& #2 \\
$n-2$& #3 \\
1& #4 \\
\cmidrule{2-4}
$\q^{(#1)}$ &$\mask{\ge 1/2n}{#5}$&$\mask{ \ge (2n-5)/2n}{#6}$&$\mask{4/15}{#7}$
\end{tabular}
}

\newcommand{\profileagentsmall}[7]{\captionlistentry{}
\centering

\begin{tabular}{r c c c}
&\multicolumn{3}{c}{\footnotesize Profile #1}\\
\# agents& a & b & c \\
1& #2 \\
$n-2$& #3 \\
1& #4 \\
\cmidrule{2-4}
$\q^{(#1)}$ &$\mask{4/15}{#5}$&$\mask{ (n-1)/n}{#6}$&$\mask{4/15}{#7}$
\end{tabular}
}

\newcommand{\profileagentsmallmod}[7]{\captionlistentry{}
\centering

\begin{tabular}{r c c c}
&\multicolumn{3}{c}{\footnotesize Profile #1}\\
\# agents& a & b & c \\
1& #2 \\
$n-2$& #3 \\
1& #4 \\
\cmidrule{2-4}
$\q^{(#1)}$ &$\mask{4/15}{#5}$&$\mask{ (n-1)/n}{#6}$&$\mask{4/15}{#7}$
\end{tabular}
}

\renewcommand{\epsilon}{\varepsilon}

\allowdisplaybreaks

\title{Optimal Budget Aggregation with Star-Shaped Preference Domains}

\setkomafont{author}{\normalfont\small}

\author{Felix Brandt\\ TUM, Germany \and Matthias Greger\\ Paris Dauphine, France \and Erel Segal-Halevi\\Ariel Univ., Israel \and Warut Suksompong\\ NUS, Singapore}

\date{}

\begin{document}

\maketitle

\begin{abstract}
We study the problem of aggregating distributions, such as budget proposals, into a collective distribution.
An ideal aggregation mechanism would be Pareto efficient, strategyproof, and fair.
Most previous work assumes that agents evaluate budgets according to the $\ell_1$ distance to their ideal budget. 
We investigate and compare different models from the larger class of \emph{star-shaped utility functions}---a multi-dimensional generalization of single-peaked preferences. 
For the case of two alternatives, we extend existing results by proving that under very general assumptions, the \emph{uniform phantom mechanism} is the only strategyproof mechanism that satisfies proportionality---a minimal notion of fairness introduced by \citet{FPPV21a}.
Moving to the case of more than two alternatives, we establish sweeping impossibilities for $\ell_1$ and $\ell_\infty$ disutilities: no mechanism satisfies efficiency, strategyproofness, and proportionality. 
We then propose a new kind of star-shaped utilities based on evaluating budgets by the \emph{ratios} of shares between a given budget and an ideal budget. 
For these utilities, efficiency, strategyproofness, and fairness become compatible. 
In particular, we prove that the mechanism that maximizes the Nash product of individual utilities is characterized by group-strategyproofness and a core-based fairness condition.
\end{abstract}

\section{Introduction}\label{sec:intro}

Social choice theory is concerned with the aggregation of individual preferences into a collective outcome \citep[e.g.,][]{ASS02a,ASS11a}.
An important special case arises when the potential collective outcomes are \emph{distributions} over a fixed set of alternatives.
These distributions may represent how a budget should be divided among public projects in a city or among departments in an organization.
Alternatively, they may reflect how time or space ought to be allotted between different types of activities at a social event.
This scenario is sometimes referred to as \emph{budget aggregation} or \emph{portioning} and falls under the framework of \emph{participatory budgeting}, which has received increasing interest in recent years \citep{AzSh21a,DNS22a}.

In order to reason about the agents' satisfaction with the collective outcome, one needs to make some assumptions about their preferences. 
Importantly, 
in our setting, the realized outcome is a distribution.
Therefore, restricting attention to rankings over alternatives is insufficient, as an agent's most preferred outcome is typically a non-degenerate distribution over the alternatives.
This is particularly evident in participatory budgeting problems, where even if an agent has a favorite project, she normally also likes other projects and does not want them to be left completely unfunded.
This is in contrast to probabilistic social choice \citep[see, e.g.,][]{Bran17a}, where the final outcome is a single alternative picked at random from the distribution, so typically the agents' most-preferred distributions are degenerate.

In this paper, 
consistent with previous research in this domain, we consider utility models where agents' preferences are completely determined by their favorite distribution: their ``peak''. This keeps the amount of required information from each agent at a manageable level.
We assume that each agent's utility decreases as the actual distribution moves away from her peak (see the formal definition in \Cref{sec:prelims}).
Such utility functions are called \emph{star-shaped} \citep{BoJo83a}.

When there are only two alternatives, the set of outcomes can be represented by the unit interval $[0,1]$. Then, star-shaped utilities are equivalent to \emph{single-peaked preferences}. 
According to a famous characterization by 
\citet{Moul80a}, there is a rather restrictive class of mechanisms that are strategyproof for all single-peaked preferences, the so-called \emph{generalized median} rules. 
\citeauthor{Moul80a}'s characterization leaves open the possibility that, for restricted subdomains of single-peaked preferences, other mechanisms than generalized median rules are strategyproof.
In \Cref{sec:m=2}, we obtain characterizations of continuous mechanisms which hold not only for single-peaked, but also for \emph{any} subdomain of single-peaked preferences. Our characterizations refine results by \citet{FPPV21a} and \citet{ALLW25a}, which are valid only for specific subdomains.

When there $m\geq 3$ alternatives, the set of outcomes is a multi-dimensional simplex, and the space of possible utility functions is much richer. For example, for every $p\geq 1$, the \emph{$\ell_p$ metric} induces a utility function 
$u_i(\q) = -\left(\sum_{j=1}^m |p_{i,j} - q_j|^p\right)^{1/p}$, where $\q = (q_1, \dots, q_m)$ is any distribution over $m$ alternatives and $\p_i = (p_{i,1},\dots,p_{i,m})$ is agent~$i$'s peak.
Note that all these functions are equivalent when $m=2$ but not when $m\geq 3$.
Still, most of the previous works on budget aggregation assume $\ell_1$ utilities  \citep[e.g.,][]{LNP08a,GKSA19a,FPPV21a,CCP24a,FrSc24a}.
Under this utility model, \citet{LNP08a} and \citet{GKSA19a} showed that a mechanism that maximizes utilitarian welfare (i.e., minimizes the sum of agents' disutilities) is both strategyproof and efficient; however, this mechanism has a tendency to overweight majority preferences.
\citet{FPPV21a} proposed a mechanism, called the \emph{independent markets mechanism}, which satisfies strategyproofness along with a weak fairness notion dubbed \emph{proportionality}. Proportionality requires that the collective distribution is given by the uniform distribution over the agents' peaks whenever all peaks are degenerate.
The independent markets mechanism violates efficiency, and \citeauthor{FPPV21a}~raised the question whether there are mechanisms that satisfy all three properties simultaneously. 
In \Cref{sec:imposs}, we settle this question by proving that no such mechanism exists under 
$\ell_1$ as well as under $\ell_\infty$ preferences. 

Using $\ell_1$ distances to define preferences over distributions has some shortcomings when aiming for a suitable representation of alternatives in the collective distribution. For instance, if the agents are deciding the amount of time that should be allotted to three countries at an international conference, an agent with an ideal distribution of $(10\%, 40\%, 50\%)$ would find the outcome $(0\%, 45\%, 55\%)$ to be quite desirable according to the $\ell_1$~distance, despite the fact that this outcome leaves the first country completely unrepresented.
Moreover, any distance-based metric that aggregates coordinate-wise differences, such as $\ell_p$ for any $p\geq 1$, has similar shortcomings. For example, if a citizen believes that the two larger districts deserve $40\%$ of the city budget each and the two smaller districts $10\%$ each, then $\ell_p$~preferences state that she is indifferent between the distributions $(50\%,30\%,10\%,10\%)$ and $(40\%,40\%,20\%,0\%)$, since for both distributions, the multiset of coordinate-wise differences is $\{0\%,0\%,10\%,10\%\}$. But intuitively, the latter distribution is worse, as it leaves the last district without any funds. 
As a consequence, a different type of utility function is necessary to capture the representation of alternatives with respect to the ideal distribution.

We introduce a new class of star-shaped utility functions to the budget aggregation setting, where each agent's utility for a distribution equals the \emph{smallest quotient}, over all alternatives, that the distribution preserves in comparison to the ideal distribution.
Formally, agent~$i$'s utility for a distribution~$\q$ is given by $\min_{j} q_j/p_{i,j}$, where the minimum is taken over all alternatives~$j$ for which $p_{i,j} > 0$. 
These utility functions are a special case of Leontief utility functions as commonly studied in economics, especially in consumer theory, with goods corresponding to alternatives \citep[see, e.g.,][]{Nico04a,LiXu13a,GHP19a}. In our setting, the total amount of goods is fixed, which allows us to interpret the relative distribution of Leontief weights as an ideal distribution. Agents then want all alternatives to receive as large a fraction as possible of their ideal amounts. We will refer to these utility functions as Leontief utility functions in the following.
Leontief utilities are arguably more suitable than $\ell_1$~preferences in applications where the representation of the alternatives is crucial---indeed, for both examples in the previous paragraph, distributions that allocate none of the budget to some alternative are least preferred among all distributions according to Leontief utilities.
Not surprisingly, mechanisms that provide desirable properties such as strategyproofness with respect to $\ell_1$~preferences may fail to do so with Leontief utilities.\footnote{Concretely, suppose that there are $n = 2$ agents and $m = 3$ alternatives.
The agents' ideal distributions are $\p_1 = (0.8,0.2,0)$ and $\p_2 = (0.8,0,0.2)$, respectively.
The independent markets mechanism of \citet{FPPV21a} returns the distribution $\q = (0.6,0.2,0.2)$.
However, if the first agent reports $\p_1' = (0.82,0.18,0)$ instead, the mechanism returns $\q' = (0.62,0.18,0.2)$, which the agent prefers to $\q$ under Leontief utilities.}
Therefore, one needs to find different mechanisms when dealing with Leontief utilities. 
In \Cref{sec:char}, we show that maximizing Nash welfare, i.e., the product of agents' utilities, results in a mechanism with several desirable properties. 
In fact, the impossibility for $\ell_1$ preferences established in \Cref{sec:imposs} can be turned into a complete characterization for Leontief utilities:
only the Nash product rule satisfies group-strategyproofness and a natural core-based fairness notion, which strengthens both efficiency and proportionality.
Thus, in contrast to $\ell_1$ preferences, Leontief utilities allow for the efficient, strategyproof, and fair aggregation of budgets via a unique attractive mechanism.

\section{Related work}\label{sec:relwork}

Various streams of research have investigated the strategyproof aggregation of preferences when the space of alternatives is a subset of a multi-dimensional Euclidean space. 

\paragraph{Public goods provision.}
One setting that calls for such an aggregation occurs when the society has to decide on the amounts of $m$ public goods.
In this context, \citet{BaPe90a} proved that any mechanism which is strategyproof for all continuous utilities, and whose range has more than two alternatives, is a dictatorship. 
Their result strongly hinges on the richness of the domain of preferences. 

\citet{Zhou91b} proved an analogous theorem for the much more restricted class of \emph{quadratic preferences}, a generalization of $\ell_2$-based preferences which is a special case of star-shaped preferences. 
Specifically, he proved that when the space of alternatives is an arbitrary convex subset of some finite-dimensional Euclidean space and the admissible domain contains all quadratic preferences, every strategyproof mechanism whose image has dimension at least $2$ is dictatorial.
\citet{BaJa94a} complemented Zhou's result by characterizing the strategyproof mechanisms whose image has dimension $1$.

\paragraph{Probabilistic social choice.}
Any model in which preferences over the set of all lotteries $\Delta^m$ are aggregated to a collective lottery, including literature on probabilistic social choice \citep[e.g.,][]{Gibb77a} and fair mixing \citep[e.g.,][]{BMS05a}, can be interpreted in the context of budget aggregation. However, the underlying assumptions of linear preferences (vNM utilities) are hardly applicable to budget aggregation. This is because,  when utility functions are linear, the maximum utility must be attained in a corner of the simplex (at least when there is a unique maximum), which corresponds to a degenerate ideal distribution.

\paragraph{Peak-based preferences.}
To the best of our knowledge, \citet{Intr73a} was the first to consider a model in which each agent has a unique ideal distribution, and the considered mechanisms only need to aggregate individual distributions into a collective distribution. He proposed three simple axioms for this setting, which characterize the rule that returns the average (i.e., the arithmetic mean) of all individual distributions. \citeauthor{Intr73a} was not concerned with strategyproofness, and therefore did not consider agents' preferences over distributions, but it is obvious that the average rule is manipulable for almost any reasonable definition of preferences.

\citet{BoJo83a} studied separable star-shaped preferences when the space of alternatives is the $m$-dimensional Euclidean space $\mathbb{R}^m$. They focused on quadratic preferences and showed that strategyproof mechanisms can be decomposed into one-dimensional mechanisms.
However, their results do not carry over to budget aggregation because their space of alternatives is the full Euclidean space and does not have a ``budget constraint''.

\paragraph{$\ell_1$-based preferences.}
\citet{LNP08a} initiated the study of strategyproof mechanisms for the space of all distributions $\Delta^m$ when preferences are based on the $\ell_1$-norm \citep[see also][]{Lind11a,GKSA19a}. They showed that the utilitarian rule (i.e., minimizing the sum of $\ell_1$ distances) satisfies strategyproofness and efficiency when breaking ties appropriately. 

\citet{FPPV21a} expanded the idea of \citeauthor{Moul80a}'s generalized median rules for two alternatives to strategyproof \emph{moving phantom mechanisms} for larger numbers of alternatives $m$. 
Intuitively, the $n+1$ phantoms are not ``fixed'' like in \citeauthor{Moul80a}'s characterization but increase continuously from $0$ to $1$ over time. For any point in time and any alternative $j$, the mechanism computes the median of $p_{1,j},\dots,p_{n,j}$ and the phantom voters. \citeauthor{FPPV21a}~then showed that there exists a well-defined point in time where the $m$ medians sum up to $1$ and thus form a valid distribution. Furthermore, they proved that within this class, maximizing utilitarian welfare is the unique efficient mechanism.
A different mechanism in this class, the \emph{independent markets} mechanism, is inefficient but satisfies a fairness notion they called \emph{proportionality}: when all voters have degenerate peaks, the collective distribution is the arithmetic mean of these peaks. \citeauthor{FPPV21a}~observed an ``inherent tradeoff between Pareto optimality and proportionality'' for strategyproof mechanisms. We prove this tradeoff formally in \Cref{thm:imposs}, which shows that all three properties are incompatible.
Proportionality was generalized by \citet{CCP24a}, who measured the ``disproportionality'' of a mechanism as the worst-case $\ell_1$ distance between the mechanism outcome and the mean. 
Similarly, \citet{FrSc24a} measured disproportionality using the $\ell_{\infty}$ distance. Both papers present variants of moving phantom mechanisms that guarantee low disproportionality.
\citet{EST23a} defined various axioms for budget aggregation with $\ell_1$ disutilities, analyzed the implications between axioms, and determined which axioms are satisfied by common aggregation rules.

\paragraph{Belief aggregation.} Belief aggregation is a setting in which several experts have different beliefs, expressed as probability density functions over a set of potential outcomes. The goal is to construct a single aggregated distribution. Technically, the problem is similar to budget aggregation; however, the utility functions are often different. \citet{VaLa22a} assumed that the outcomes are linearly ordered (for example: outcome $j$ is an earthquake of magnitude $j$). Then, an expert whose belief is $3$ with probability $1$ would prefer the outcome $6$ with probability $1$ to the outcome~$9$ with probability $1$, even though the $\ell_1$ distance is $2$ in both cases. \citeauthor{VaLa22a} suggested preferences based on distance between the \emph{cumulative} distribution functions, and characterized aggregation rules satisfying appropriate strategyproofness and proportionality axioms.

\paragraph{Donor Coordination.} \citet{BGSS23a} studied \emph{donor coordination}, where individual monetary contributions by agents are distributed on projects based on the agents' preferences. 
Assuming Leontief preferences, they proposed the \emph{equilibrium distribution rule (\edr)}, which maximizes Nash welfare and distributes the contributions of the donors in such a way that no subset of donors has an incentive to redistribute their contributions.
\edr can be interpreted as a budget aggregation mechanism, by setting the contribution of each agent to $1/n$ (where $n$ is the number of agents, so the total contribution is~$1$) and setting the ideal distribution of an agent to the distribution given by the relative proportions of the Leontief weights.
This allows the transfer of positive results concerning efficiency and strategyproofness of \edr from donor coordination to budget aggregation (see \Cref{sec:char} and Appendix \ref{sec:nashproofs}). 
However, other properties considered by \citeauthor{BGSS23a}, like contribution-monotonicity and being in equilibrium, are irrelevant in budget aggregation, whereas properties like core fair share and proportionality have not been considered in donor coordination.

\section{Preliminaries}\label{sec:prelims}

Let $N = [n]$ be a set of agents and $M = [m]$ be a set of alternatives, where $[k] \coloneqq \{1,\dots,k\}$ for each positive integer~$k$.
We denote by $\Delta^m$ the standard simplex with $m$ vertices, that is, the set of vectors $\q = (q_1,\dots,q_m)$ with nonnegative entries $q_j$ such that $\sum_{j\in M} q_j = 1$.
Every element $\q\in\Delta^m$ is called a \emph{distribution}; here, $q_j$ denotes the fraction of a public resource (e.g., money or time) allocated to alternative~$j$.
The support of distribution $\q$ is given by $\supp(\q)\coloneqq\{j\in M\colon q_j>0\}$.
For a set of alternatives $T\subseteq M$, we denote
$\q(T) \coloneqq \sum_{j \in T}q_j$.

Each agent $i$ has a utility function $u_i$ over distributions; we denote by $\mathcal{U}$ the set of all possible utility functions and leave this set unspecified for now.

The \emph{ideal distribution} or \emph{peak} of agent $i$ is denoted by $\p_i \coloneqq \argmax_{\q \in \Delta^m}u_i(\q)$, which is assumed to be unique.
We further assume that ``walking'' towards an agent's peak strictly increases her utility. Formally, if agent $i$ has peak $\p_i$ and utility function $u_i$, then, for any distribution $\q \neq \p_i$,
the agent prefers to $\q$ any distribution on the line between $\q$ and $\p_i$. That is, for all $\lambda \in (0,1)$,
\begin{align}
\label{eq:star-shaped}
u_i(\p_i)>u_i(\lambda \p_i+(1-\lambda)\q)>u_i(\q)\text. 
\end{align}

This constitutes a generalization of single-peakedness \citep{Blac48a} and is referred to as \emph{star-shaped} or star-convex preferences \citep[e.g.,][]{BoJo83a}.%
\footnote{
\citet{BOS20a} use a weaker definition to star-shaped preferences: for every distribution $\q$, the set of distributions that are weakly-preferred to $\q$ is a star domain. This definition is implied by the definition we use, as for any distribution $\q$, if $\q'$ is preferred to $\q$, then by transitivity, all distributions on the line from $\p_i$ to $\q'$ are also preferred to $\q$, so the set of distributions weakly-preferred to~$\q$ is a star domain with respect to $\p_i$.
Their definition is weaker as it allows the domain to be a star with respect to another point.
}

The utility functions considered in this paper belong to a subclass of star-shaped preferences that we refer to as \emph{peak-linear}. A utility function is peak-linear if for any distribution $\q$ and $\lambda \in [0,1]$,
\begin{align}
\label{eq:peak-linear}
u_i(\lambda \p_i+(1-\lambda)\q)=\lambda u_i(\p_i) + (1-\lambda) u_i(\q)\text. 
\end{align}
Every distribution $\q \in \Delta^m$ lies on a line between $\p_i$ and some point $\q^B$ on the boundary of $\Delta^m$, that is, there is a unique $\lambda$ such that $\q = \lambda \p_i+(1-\lambda)\q^B$. Therefore, peak-linear preferences are completely characterized by an agent's peak $\p_i$ and by how much utility she assigns to these boundary distributions $\q^B$
(in fact, it is sufficient to know how much utility she assigns to boundary distributions with $\supp(\p_i)\not\subseteq \supp(\q^B)$). 

Another important property of preferences is \emph{convexity}. 
An agent has convex preferences if for every three outcomes $\q$, $\q'$, and $\q''$ where $\q'$ and $\q''$ are both weakly preferred to $\q$, any mixture $\lambda \q'+(1-\lambda)\q''$ with $\lambda \in [0,1]$ is also weakly preferred to $\q$. For \emph{strict convexity}, any mixture of $\q'$ and $\q''$ with $\lambda \in (0,1)$ has to be strictly preferred to $\q$ unless $\q'=\q''$ \citep[see, e.g.,][]{MWG95a}.
All strictly convex preferences are star-shaped, but this does not hold for all convex preferences: assuming only convexity is neither sufficient to guarantee a unique peak, nor does it imply that an agent's utility is \emph{strictly} increasing when moving towards her peak.
The preferences we consider here (specifically, $\ell_1, \ell_{\infty}$, and Leontief) are star-shaped and convex, but not strictly convex.
For example, Leontief preferences indicate indifference between all distributions $\q$ with $\supp(\p_i)\not\subseteq \supp(\q)$. Therefore, we use star-shaped/peak-linear preferences as a collective term for all investigated utility models.

Unless explicitly stated otherwise, we further assume that $\p_i$ completely determines~$u_i$,
so we can identify a profile $P \in \mathcal{P}\coloneqq(\Delta^m)^n$ 
with the matrix $(p_{i,j})_{i \in N,j \in M}$ containing the peaks $\p_1,\dots,\p_n$ as rows.

A \emph{mechanism} $f: \mathcal{P} \to \Delta^m$ aggregates individual peaks into a collective distribution. In the following, we define desirable properties of aggregated distributions and mechanisms.

\subsection{Properties of distributions}

Two important properties of distributions are efficiency and fairness.
We consider two efficiency properties.

\begin{definition}
    A distribution $\q \in \Delta^m$ satisfies \emph{(Pareto) efficiency} if  there does not exist a distribution $\q' \in \Delta^m$ such that $u_i(\q') \ge u_i(\q)$ for all $i \in N$ and $u_i(\q') > u_i(\q)$ for at least one $i \in N$. 
\end{definition}

\begin{definition}
     A distribution  $\q \in \Delta^m$ is \emph{range-respecting} for profile $P$ if $\min_{i \in N} p_{i,j} \le q_j \le \max_{i \in N} p_{i,j}$ for all $j\in M$.
\end{definition}

With $\ell_1$ utilities, range-respect is weaker than Pareto-efficiency \citep[see, e.g.,][]{FPPV21a,EST23a}.

Our first fairness axiom is inspired by the core in cooperative game theory and was transferred to participatory budgeting by \citet{FGM16a} and to fair mixing by \citet{ABM20a} under the name of core fair share. We adapt the definition to account for the fact that, in the end, we still need to choose a probability distribution $\p$ (and not just a partial distribution $(|N'|/n)\p$) and use the weak core rather than the strong one. In fact, this leads us back to the original definition of the \emph{$\alpha$-core}, due to \citet{Auma61b} and \citet{Scar71a}.\footnote{For Leontief preferences, the weak and the strong version of the core are equivalent (see \Cref{rem:weakcfsLeon}). For dichotomous utilities, our definition is weaker than that of \citet{ABM20a} and thus their positive result about the Nash product rule carries over. 
\citet{Scar71a} has shown the non-emptiness of the weak $\alpha$-core under very general conditions. By contrast, the strong $\alpha$-core is already empty for dichotomous utilities.}

\begin{definition}
A group of agents $N' \subseteq N$ is called a \emph{blocking coalition} for a distribution~$\q$ if there exists a distribution $\q'\in \Delta^m$ such that for every $\q'' \in \Delta^m$,
\begin{align*}
u_i((|N'|/n)\q'+(1-|N'|/n)\q'') > u_i(\q) && \text{for all $i \in N'$.}
\end{align*}

A distribution $\q$ satisfies \emph{core fair share} if it has no blocking coalition.
\end{definition}

Intuitively, a blocking coalition $N'$ can take their share of the decision power ($|N'|/n$), and redistribute it via $\q'$ so that all members of~$N'$ gain utility compared to $\q$, even if the remainder is distributed in the worst possible way $\q''$ (e.g., on an alternative that no agent from $N'$ values).

We consider another, weaker fairness axiom that is only informative on specific profiles in the next subsection.

\subsection{Properties of aggregation mechanisms}

\begin{definition}
A mechanism $f$ satisfies efficiency (resp., core fair share) if for every profile $P \in \mathcal{P}$, $f(P)$ satisfies efficiency (resp., core fair share). 
\end{definition}

The next axioms ensure that agents and alternatives are treated independently of their identities.

\begin{definition}
    A mechanism $f$ satisfies \emph{anonymity} if for every profile $P \in \mathcal{P}$ and permutation~$\pi$ of the agents in $P$, it holds that $f(P)=f(\pi \circ P)$.
\end{definition}

\begin{definition}
    A mechanism $f$ satisfies \emph{neutrality} if for every profile $P \in \mathcal{P}$ and permutation $\pi$ of the alternatives resulting in profile $P'$, it holds that $f(P')=\pi \circ f(P)$.
\end{definition}

As agents report a peak in $\Delta^m$, we do not want small perturbations of the peaks arising from uncertainties of the agents about their exact peak or inaccuracies during the aggregation process to have a large influence on the outcome. 

\begin{definition}
    A mechanism $f$ satisfies \emph{continuity} if 
    \begin{align*}
        \forall P \in \mathcal{P}:\; \forall \epsilon>0:\; \exists \delta>0:\; \forall P' \in \mathcal{P}: \|P-P'\|_1<\delta \implies \|f(P)-f(P')\|_1<\epsilon.
    \end{align*}
\end{definition}
For simplicity, we define continuity using the $\ell_1$ distance, but note that due to the norm equivalence on finite-dimensional vector spaces, 
our results are the same for every other norm-based distance.

As $\mathcal{P}=(\Delta^m)^n$ 
is compact with respect to the $\ell_1$ distance (or other equivalent norms), the Heine-Cantor theorem implies that a continuous mechanism $f$ is also \emph{uniformly continuous}, i.e.,
\begin{align*}
    \forall \epsilon>0: \;\exists \delta>0: \; \forall P,P' \in \mathcal{P}: \|P-P'\|_1<\delta \implies \|f(P)-f(P')\|_1<\epsilon.
\end{align*}
This insight will play an important role in the proof of \Cref{thm:Nashchar}.

Another common goal is to prevent agents from misreporting their peaks on purpose.

\begin{definition}
    A mechanism $f$ satisfies \emph{group-strategyproofness} if for all $N'\subseteq N$ and all $P,P' \in \mathcal{P}$ with $\p_j=\p'_j$ for all $j \not \in N'$, either $u_i(f(P)) > u_i(f(P'))$ for at least one $i \in N'$ or $u_i(f(P)) = u_i(f(P'))$ for all $i \in N'$, where $u_i$ refers to the utility function of agent~$i$ with peak at $\p_i$.
    The mechanism $f$ satisfies \emph{strategyproofness} if the above statement holds for $|N'|=1$. 
\end{definition}

Finally, we consider another fairness property called \emph{proportionality} by \citet{FPPV21a}. 
It restricts the set of outcomes only on profiles where all agents are ``single-minded'', thus representing a rather weak form of ``traditional'' proportionality considered, e.g., in fair division.

\begin{definition}
    A profile $P \in \mathcal{P}$ is called \emph{single-minded} if 
    $p_{i,j} \in \{0,1\}$ for all $i \in N$ and $j \in M$.
\end{definition}

\begin{definition}
    A mechanism $f$ satisfies \emph{proportionality} if for all single-minded profiles $P \in \mathcal{P}$, it holds that $f(P)_j=\sum_{i \in N}p_{i,j}/n$.
\end{definition}

The following diagram shows logical relationships between efficiency and the fairness notions we consider.
The proofs of Propositions~\ref{prop:Leontief+cfs->eff} and~\ref{prop:cfs-implies-prop} can be found in Appendix~\ref{sec:Leontiefcollection}.

\medskip

\begin{center}
\begin{tikzcd}[column sep=10em]
 \hspace{-0.2em}\text{Efficiency}\hspace{-1em}
\ar[r,shorten >=1em,shorten <=1em, "\text{\color{gray}\Cref{prop:Leontief+cfs->eff}  (Leontief)}",Latex-]
& \hspace{-1em}\text{Core fair share}\hspace{0.4em}
    \ar[r,shorten >=-0.5em, shorten <=-0.4em, "{\text{\color{gray}\Cref{prop:cfs-implies-prop}  ($\ell_1$, $\ell_\infty$, Leontief)}}",-Latex]
& \hspace{0.3em}\text{Proportionality}
\end{tikzcd}
\end{center}
\medskip

\section{Specific star-shaped preference domains}
\label{sec:ufunctions}

We initially discuss some typical star-shaped utility functions, for which we will present results in the next sections.

\subsection{$\ell_p$ preferences}
A natural approach for specifying a utility function based on a single peak is to measure the distance to the peak using some metric $d:\Delta^m \times \Delta^m \to \mathbb{R}_{\geq 0}$. Given an agent with peak $\p_i$, her utility for a distribution $\q$ is then defined as $u_i(\q)=-d(\p_i,\q)$. 

Among such models, $\ell_p$ norms, given by $ \|\q\|_p \coloneqq \left(\sum_{j\in M} |q_j|^p\right)^{1/p}$ for $p \geq 1$, are the most studied utility functions. In particular, the special case of $p=1$ has received considerable attention.

\begin{definition}
    An agent $i$ with peak $\p_i$ has \emph{$\ell_p$ preferences} if $u_i(\q)=-\|\p_i-\q\|_p$.
\end{definition}
We will also sometimes refer to these preferences as $\ell_p$ disutilities. In this paper, we focus on $\ell_1$ preferences ($u_i(\q)=-\sum_{j \in M}|p_{i,j}-q_j|$) and $\ell_\infty$ preferences ($u_i(\q)=- \max_{j \in M}|p_{i,j}-q_j|$).
It can be easily shown that $\ell_p$ preferences are star-shaped for $p \geq 1$. 
This does not hold for arbitrary metrics, e.g., consider the \emph{trivial metric}, $d(x,y)=0$ if $x=y$ and $d(x,y)=1$ otherwise. 
Moreover, $\ell_p$ preferences are peak-linear. The utility of every point in $\Delta^m$ can be computed using \eqref{eq:peak-linear}
with $u_i(\p_i)=0$ and $u_i(\q)=-\| \p_i-\q\|_p$ for all distributions $\q$ on the boundary of $\Delta^m$.

\subsection{Leontief utilities}
\label{sub:mq-utils}

In contrast to $\ell_p$ preferences, Leontief utilities are not based on a metric. In particular, they are not symmetric and also not based on disutilities.
As discussed in the introduction, metric-based preferences fail to capture important aspects of certain practical situations, notably the need to guarantee that all alternatives are adequately represented. This requirement is captured by Leontief utilities.

Let $M_i \coloneqq \{j\in M\colon p_{i,j} > 0\}$ be the set of alternatives to which $i$ wants to allocate a positive amount; note that $M_i\ne\emptyset$.
The \emph{Leontief utility}  that agent~$i$ derives from a distribution $\q$ is given by 
\[
u_i(\q) = \min_{j\in M_i}\frac{q_j}{p_{i,j}}.
\]
Observe that $0\le u_i(\q)\le 1$ for all distributions~$\q$.
Moreover, $u_i(\q) = 1$ if and only if $\q = \p_i$, and $u_i(\q) = 0$ if and only if $q_j = 0$ for some $j\in M_i$.
As discussed in \Cref{sec:intro}, Leontief utilities are based on the assumption that agents want all alternatives to receive as large a fraction of their ideal amounts as possible.
The indifference curves of $\ell_1$ preferences and Leontief utilities are illustrated in \Cref{fig:indifference}.

\begin{figure}
\centering

\begin{subfigure}{.33\textwidth}
\centering
\begin{tikzpicture}
\coordinate (a) at (0,0);
\coordinate (b) at (4,0);
\coordinate (c) at (2,{sqrt(12)});
\draw (a.north east) -- (b.north west) -- (c.south) -- cycle;
\fill (barycentric cs:a=0.1,b=0.4,c=0.5) circle (2pt);
\coordinate (a1) at (barycentric cs:a=26/80,b=3/10,c=3/8);
\coordinate (b1) at (barycentric cs:a=3/40,b=22/40,c=3/8);
\coordinate (c1) at (barycentric cs:a=3/40,b=3/10,c=25/40);
\draw[ultra thick] (a1) -- (b1) -- (c1) -- cycle;
\coordinate (a2) at (barycentric cs:a=11/20,b=1/5,c=1/4);
\coordinate (b2) at (barycentric cs:a=1/20,b=14/20,c=1/4);
\coordinate (c2) at (barycentric cs:a=1/20,b=1/5,c=15/20);
\draw[very thick] (a2) -- (b2) -- (c2) -- cycle;
\coordinate (a3) at (barycentric cs:a=31/40,b=1/10,c=1/8);
\coordinate (b3) at (barycentric cs:a=1/40,b=34/40,c=1/8);
\coordinate (c3) at (barycentric cs:a=1/40,b=1/10,c=35/40);
\draw[thick] (a3) -- (b3) -- (c3) -- cycle;
\end{tikzpicture}
\caption{Leontief utilities}
\end{subfigure}
\qquad\qquad\qquad
\begin{subfigure}{.33\textwidth}
\centering
\begin{tikzpicture}
\coordinate (a) at (0,0);
\coordinate (b) at (4,0);
\coordinate (c) at (2,{sqrt(12)});
\draw (a.north east) -- (b.north west) -- (c.south) -- cycle;
\fill (barycentric cs:a=0.1,b=0.4,c=0.5) circle (2pt);
\coordinate (ab1) at (barycentric cs:a=0.1+1/6,b=0.4,c=0.5-1/6);
\coordinate (ac1) at (barycentric cs:a=0.1+1/6,b=0.4-1/6,c=0.5);
\coordinate (ba1) at (barycentric cs:a=0.1,b=0.4+1/6,c=0.5-1/6);
\coordinate (bc1) at (barycentric cs:a=0,b=0.4+1/6,c=0.5+0.1-1/6);
\coordinate (ca1) at (barycentric cs:a=0.1,b=0.4-1/6,c=0.5+1/6);
\coordinate (cb1) at (barycentric cs:a=0,b=0.4+0.1-1/6,c=0.5+1/6);
\draw[ultra thick] (bc1) -- (ba1) -- (ab1) -- (ac1) -- (ca1) -- (cb1);
\coordinate (ab2) at (barycentric cs:a=0.1+1/3,b=0.4,c=0.5-1/3);
\coordinate (ac2) at (barycentric cs:a=0.1+1/3,b=0.4-1/3,c=0.5);
\coordinate (ba2) at (barycentric cs:a=0.1,b=0.4+1/3,c=0.5-1/3);
\coordinate (bc2) at (barycentric cs:a=0,b=0.4+1/3,c=0.5+0.1-1/3);
\coordinate (ca2) at (barycentric cs:a=0.1,b=0.4-1/3,c=0.5+1/3);
\coordinate (cb2) at (barycentric cs:a=0,b=0.4+0.1-1/3,c=0.5+1/3);
\draw[very thick] (bc2) -- (ba2) -- (ab2) -- (ac2) -- (ca2) -- (cb2);
\coordinate (ab3) at (barycentric cs:a=0.6,b=0.4,c=0);
\coordinate (ac3) at (barycentric cs:a=0.6,b=0,c=0.4);
\coordinate (ba3) at (barycentric cs:a=0.1,b=0.9,c=0);
\coordinate (bc3) at (barycentric cs:a=0,b=0.9,c=0.1);
\coordinate (c3) at (barycentric cs:a=0,b=0,c=1);
\draw[thick] (bc3) -- (ba3) -- (ab3) -- (ac3);
\fill (barycentric cs:a=0,b=0,c=1) circle (0.7pt);
\end{tikzpicture}
\caption{$\ell_1$ disutilities}
\end{subfigure}
\label{fig:indifference}
\caption{Illustration of indifference classes for Leontief utilities and $\ell_1$~disutilities for $3$ alternatives.
The ideal distribution $(0.1,0.4,0.5)$ is represented by the black point.
The main triangle is the simplex of distributions among three alternatives. 
Its vertices represent the degenerate distributions $(1,0,0)$, $(0,1,0)$, and $(0,0,1)$.
For each type of utilities, the peak forms an indifference class by itself, and three other indifference classes are displayed with different line widths. 
}
\end{figure}

Leontief utility functions are peak-linear with $u_i(\p_i)=1$ and $u_i(\q)=0$ for all boundary distributions $\q$. 
In fact, as explained in the preliminaries, Leontief utilities are characterized by these properties: they are the only peak-linear utilities that assign utility $1$ to the peak and utility $0$ to all boundary distributions.
Some useful concepts and insights on Leontief utilities are collected in Appendix \ref{sec:Leontiefcollection}. 

It is possible to refine Leontief utilities further by considering the \emph{leximin} over the quotients---that is, breaking ties in the smallest quotient using the second smallest quotient, and so on. This refinement is discussed in Appendix~\ref{sec:leximin}.

\section{Impossibilities for $\ell_1$ and $\ell_{\infty}$ preferences}\label{sec:imposs}

In this section, we show that efficiency, strategyproofness, and the rather weak fairness condition of proportionality (cf.~\Cref{prop:cfs-implies-prop}) are incompatible when agents have $\ell_1$ or $\ell_{\infty}$ preferences.

\subsection{$\ell_1$ preferences}

Under $\ell_1$ preferences, \citet{FPPV21a} observed that the utilitarian welfare maximizing mechanism is the only efficient mechanism in their class of moving phantom mechanisms. However, maximizing utilitarian welfare violates weak fairness axioms such as proportionality. We prove that this tradeoff between efficiency and fairness is inevitable in the presence of strategyproofness.

\begin{theorem}\label{thm:imposs}
    With $\ell_1$ preferences, no mechanism satisfies efficiency, strategyproofness, and proportionality when $m \ge 3$ and $n \ge 3$.
\end{theorem}

For the proof of this theorem, we consider disutilities $d_i$ (the $\ell_1$ distance to an agent's peak~$p_i$) instead of utilities, i.e., $d_i(\q)=\|\p_i-\q\|_1$. It is also important to keep the following simple observations in mind.

\begin{observation}
\label{obs:ell1}
With $\ell_1$ preferences, if some agent $i$ has $p_{i,j}=1$, then for any distribution $\q$, $d_i(\q)=2-2q_j$, regardless of the distribution on alternatives other than~$j$. Therefore, agent $i$ is indifferent if some amount is moved between alternatives other than~$j$.
\end{observation}

\begin{observation}
\label{obs:ell1two}  
With $\ell_1$ preferences, $d_i(\q) \ge 2 \cdot \lvert p_{i,j}-q_j\rvert$ for all $j \in M$ and $i \in N$.
\end{observation}

\begin{proof}[Proof of \Cref{thm:imposs}]
We start with the case $m=3$ and $n=3$.
For $m=3$, we set $M=\{a,b,c\}$ and write $\q=(q_a,q_b,q_c)$. 
For simplicity, we number profiles by a superscript~$(k)$. We denote the disutility function of agent $i$ in profile $k$ by $d^{(k)}_i$, and the returned distribution in profile $k$ by $\q^{(k)}$. Sometimes, we cannot determine $\q^{(k)}$ completely. In these cases, we give lower or upper bounds on the entries of $\q^{(k)}$.

Consider first the following two profiles.
The outcome in Profile \ref{prof:2l1} must be $(1/3,1/3,1/3)$ by \prop.
For Profile \ref{prof:1l1}, the last row of the table gives lower bounds on $\q^{(1)}_a,\q^{(1)}_b$ and an upper bound on $\q^{(1)}_c$, which we justify below.

\begin{table}[h]
\parbox{.49\linewidth}{
    \profile{1}
      {1/2 & 1/2 & 0}{0 & 1 & 0}{0 & 0 & 1}
      {{\ge} 1/6}{{\ge} 1/2}{{\le} 1/3}
      \label{prof:1l1}
}
\parbox{.49\linewidth}{
    \profile{2}
      {1 & 0 & 0}{0 & 1 & 0}{0 & 0 & 1}
      {1/3}{1/3}{1/3}
      \label{prof:2l1}
}
\end{table}

As Agent 1 can manipulate between Profile \ref{prof:1l1} and Profile \ref{prof:2l1}, \strp requires that Agent~1 does not gain from either manipulation. This implies 
\begin{align}
    d_1^{(1)}(\q^{(1)}) &\leq d_1^{(1)}(\q^{(2)}) = 2/3 \text{ and} \label{ineq:imposs1}
    \\
    d_1^{(2)}(\q^{(1)}) &\geq d_1^{(2)}(\q^{(2)}) = 4/3. \label{ineq:imposs2}
\end{align}

By (\ref{ineq:imposs1})
and \Cref{obs:ell1two}, $q^{(1)}_a \ge 1/6$ (implying $q^{(1)}_b \le 5/6$), $q^{(1)}_b \ge 1/6$, and $q^{(1)}_c \le 1/3$.
By (\ref{ineq:imposs2}) and \Cref{obs:ell1}, $q^{(1)}_a \le 1/3$, implying $q^{(1)}_b+q^{(1)}_c \ge 2/3$, and thus $q^{(1)}_b \ge 1/3$.

\Cref{fig:3456}(a) illustrates both inequalities. The blue area corresponds to the set of distributions $\q^{(1)}$ with $d_1^{(1)}(\q^{(1)}) \leq 2/3$ whereas the red area consists of all distributions $\q^{(1)}$ satisfying (\ref{ineq:imposs2}). Strategyproofness requires $\q^{(1)}$ to be inside the intersection of the two areas, i.e., the purple region.
Hence, 
\begin{align*}
1/6\leq q^{(1)}_a \leq 1/3\text{,} 
&&
1/3\leq q^{(1)}_b \leq 5/6\text{,\qquad and}
&&
0\leq q^{(1)}_c \leq 1/3\text.
\end{align*}
By \eff, we can even show that $q^{(1)}_b \ge 1/2$. Otherwise, as $q^{(1)}_a > 0$, some small amount could be moved from $a$ to $b$. Agent $3$ is indifferent due to Observation \ref{obs:ell1} and Agent~$2$ strictly gains. Furthermore, this does not change Agent $1$'s disutility as $q^{(1)}_b<1/2$.

\begin{figure}[b]
\centering
\begin{subfigure}{.33\textwidth}
\centering
\begin{tikzpicture}
\hspace{-2mm}
\coordinate [label=below left:{$a$}] (a)   at (0,0);
\coordinate [label=below right:{$b$}] (b) at (4,0);
\coordinate [label=above:{$c$}] (c) at (2,{sqrt(12)});
\draw (a.north east) -- (b.north west) -- (c.south) -- cycle;
\coordinate [label=below:{\textcolor{blue}{$\p^{(1)}_1$}}] (p1) at (barycentric cs:a=0.5,b=0.5,c=0);
\draw[fill=blue] (p1) circle (1pt);
\coordinate (p1ab) at (barycentric cs:a=0.5-1/3,b=0.5+1/3,c=0);
\coordinate (p1ba) at (barycentric cs:a=0.5+1/3,b=0.5-1/3,c=0);
\coordinate (p1bc) at (barycentric cs:a=0.5,b=0.5-1/3,c=1/3);
\coordinate (p1ac) at (barycentric cs:a=0.5-1/3,b=0.5,c=1/3);
\draw[fill=blue,opacity=0.2] (p1ba) -- (p1bc) -- (p1ac) -- (p1ab) -- (p1ba);
\coordinate [label=above left:{\textcolor{red}{$\p^{(2)}_1$}}] (p2) at (barycentric cs:a=1,b=0,c=0);
\draw[fill=red] (p2) circle (1pt);
\coordinate (p2ab) at (barycentric cs:a=1/3,b=2/3,c=0);
\coordinate (p2ac) at (barycentric cs:a=1/3,b=0,c=2/3);
\draw[fill=red,opacity=0.2] (p2ab) -- (b) -- (c) -- (p2ac) -- (p2ab);
\end{tikzpicture}
\caption{Inequalities (\ref{ineq:imposs1}) and (\ref{ineq:imposs2})}
\end{subfigure}
\qquad\qquad\qquad
\begin{subfigure}{.33\textwidth}
\centering
\begin{tikzpicture}
\hspace{3mm}
\coordinate [label=below left:{$a$}] (a)   at (0,0);
\coordinate [label=below right:{$b$}] (b) at (4,0);
\coordinate [label=above:{$c$}] (c) at (2,{sqrt(12)});
\draw (a.north east) -- (b.north west) -- (c.south) -- cycle;
\coordinate [label=below:{\textcolor{blue}{$\p^{(3)}_1$}}] (p3) at (barycentric cs:a=0.25,b=0.75,c=0);
\draw[fill=blue] (p3) circle (1pt);
\coordinate (p3ba) at (barycentric cs:a=0.25+1/3,b=0.75-1/3,c=0);
\coordinate (p3abc) at (barycentric cs:a=0,b=2/3,c=1/3);
\coordinate (p3bc) at (barycentric cs:a=1/4,b=5/12,c=1/3);
\coordinate (p3ac) at (barycentric cs:a=0.5-1/3,b=0.5,c=1/3);
\draw[fill=blue,opacity=0.2] (p3ba) -- (b) -- (p3abc) -- (p3bc) -- (p3ba);
\coordinate [label=above right:{\textcolor{red}{$\p^{(4)}_1$}}] (p4) at (barycentric cs:a=0,b=1,c=0);
\draw[fill=red] (p4) circle (1pt);
\coordinate (p4ab) at (barycentric cs:a=1/3,b=2/3,c=0);
\coordinate (p4bc) at (barycentric cs:a=0,b=2/3,c=1/3);
\draw[fill=red,opacity=0.2] (a) -- (p4ab) -- (p4bc) -- (c) -- (a);
\end{tikzpicture}
\caption{Inequalities (\ref{ineq:imposs3}) and (\ref{ineq:imposs4})}
\end{subfigure}
\caption{Illustration for the proof of \Cref{thm:imposs}.}
\label{fig:3456}
\end{figure}
Next, we consider the following two profiles.

        \begin{table}[!h]
            \parbox{.49\linewidth}{
              \profile{3}
                {1/4 & 3/4 & 0}{0 & 1 & 0}{0 & 0 & 1}
                {0}{2/3}{1/3}
                \label{prof:3l1}
            }
            \parbox{.49\linewidth}{
              \profile{4}
                {0 & 1 & 0}{0 & 1 & 0}{0 & 0 & 1}
                {0}{2/3}{1/3}
                \label{prof:4l1}
            }
            \end{table}            
The outcome in Profile \ref{prof:4l1} follows from \prop.
We now prove that the outcome in Profile~\ref{prof:3l1} must be identical. 
As Agent $1$ can manipulate between Profile \ref{prof:3l1} and Profile~\ref{prof:4l1}, \strp requires that Agent 1 does not gain from either manipulation. This implies that
\begin{align}
    d_1^{(3)}(\q^{(3)}) &\leq d_1^{(3)}(\q^{(4)}) = 2/3\text{, and} \label{ineq:imposs3}
    \\
    d_1^{(4)}(\q^{(3)}) &\geq d_1^{(4)}(\q^{(4)}) = 2/3\text. \label{ineq:imposs4}
\end{align}

By (\ref{ineq:imposs3}), $q^{(3)}_c \le 1/3$, implying $q^{(3)}_a+q^{(3)}_b \ge 2/3$.
By (\ref{ineq:imposs4}), $q^{(3)}_b \le 2/3$.
Graphically, \strp for Agent 1 implies that $\q^{(3)}$ must be in the purple region in 
\Cref{fig:3456}(b).

However, by \eff, if $q^{(3)}_a>0$ then $q^{(3)}_b\geq 3/4$. 
Otherwise, some small amount can be moved from $a$ to $b$. 
Agent $3$ is indifferent due to Observation \ref{obs:ell1} and Agent $2$ strictly gains. Furthermore, this does not change agent $1$'s disutility as $q^{(3)}_b<3/4$.
Therefore, $q^{(3)}_a=0$ must hold, and the only outcome compatible with \strp is $\q^{(3)} = (0, 2/3, 1/3)$.

Now that we know $\q^{(3)}$, we 
consider a manipulation of Agent $1$ from Profile~\ref{prof:3l1} to Profile~\ref{prof:1l1}.
Strategyproofness implies 
\begin{align*}
   d_1^{(3)}(\q^{(1)}) \geq d_1^{(3)}(\q^{(3)}) = 2/3. 
\end{align*}
But the bounds we already have for $\q^{(1)}$ 
imply that $d_1^{(3)}(\q^{(1)}) \leq 2/3$ as $q^{(1)}_a \ge 1/6$ and $q^{(1)}_b \ge 1/2$. Therefore, $d_1^{(3)}(\q^{(1)}) = 2/3$ together with $q^{(1)}_a = 1/6$ and $q^{(1)}_b=1/2$.
Hence, $\q^{(1)} = (1/6, 1/2, 1/3)$.    

Finally, we consider Profiles \ref{prof:5l1} and \ref{prof:6l1}.

        \begin{table}[h]
            \parbox{.49\linewidth}{
                \profile{5}
                  {1/2 & 1/2 & 0}{0 & 1 & 0}{0 & 1/2 & 1/2}
                  {}{}{}
                  \label{prof:5l1}
                }
            \parbox{.49\linewidth}{
                \profile{6}
                  {1 & 0 & 0}{0 & 1 & 0}{0 & 1/2 & 1/2}
                  {1/3}{1/2}{1/6}
                  \label{prof:6l1}
                }
            \end{table}

The distribution $\q^{(6)}$ is determined by arguments analogous to those for $\q^{(1)}$, reasoning about Agent 3 instead of Agent 1.

We now consider a manipulation of Agent 1 from Profile \ref{prof:5l1} to Profile \ref{prof:6l1}.
It follows from \strp that $ d_1^{(5)}(\q^{(5)})\leq d_1^{(5)}(\q^{(6)})=1/3$, which implies that $q^{(5)}_c \le 1/6$.
Similarly, we consider a manipulation of Agent 3 from Profile \ref{prof:5l1} to Profile \ref{prof:1l1}. 
It follows from \strp that $d_3^{(5)}(\q^{(5)})\leq d_3^{(5)}(\q^{(1)})=1/3$, which implies that $q^{(5)}_c \ge 1/2-1/6=1/3$, a contradiction.

\parbox{0.63\textwidth}{
Graphically, both inequalities are shown in the figure on the right.
The blue area at the bottom contains the points satisfying the first inequality, and the red area on the right contains the points satisfying the second inequality. It is evident that the two inequalities cannot be satisfied simultaneously.
}
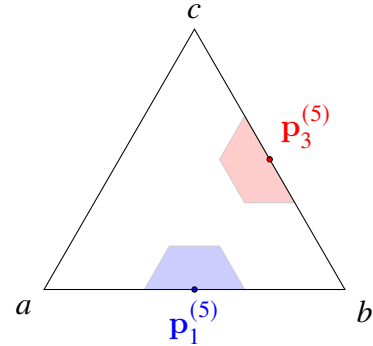
\parbox{0.35\textwidth}{
\begin{tikzpicture}
\coordinate [label=below left:{$a$}] (a)   at (0,0);
\coordinate [label=below right:{$b$}] (b) at (4,0);
\coordinate [label=above:{$c$}] (c) at (2,{sqrt(12)});
\draw (a.north east) -- (b.north west) -- (c.south) -- cycle;
\coordinate [label=below:{\textcolor{blue}{$\p^{(5)}_1$}}] (p1) at (barycentric cs:a=0.5,b=0.5,c=0);
\draw[fill=blue] (p1) circle (1pt);
\coordinate (p1ab) at (barycentric cs:a=0.5-1/6,b=0.5+1/6,c=0);
\coordinate (p1ba) at (barycentric cs:a=0.5+1/6,b=0.5-1/6,c=0);
\coordinate (p1bc) at (barycentric cs:a=0.5,b=0.5-1/6,c=1/6);
\coordinate (p1ac) at (barycentric cs:a=0.5-1/6,b=0.5,c=1/6);
\draw[fill=blue,opacity=0.2] (p1ba) -- (p1bc) -- (p1ac) -- (p1ab) -- (p1ba);
\coordinate [label=above right:{\textcolor{red}{$\p^{(5)}_3$}}] (p2) at (barycentric cs:a=0,b=0.5,c=0.5);
\draw[fill=red] (p2) circle (1pt);
\coordinate (p5cb) at (barycentric cs:a=0,b=0.5+1/6,c=0.5-1/6);
\coordinate (p5bc) at (barycentric cs:a=0,b=0.5-1/6,c=0.5+1/6);
\coordinate (p5ba) at (barycentric cs:a=1/6,b=0.5-1/6,c=0.5);
\coordinate (p5ca) at (barycentric cs:a=1/6,b=0.5,c=0.5-1/6);
\draw[fill=red,opacity=0.2] (p5bc) -- (p5ba) -- (p5ca) -- (p5cb) -- (p5bc);
\end{tikzpicture}
}

This example can be extended to arbitrary numbers of alternatives and agents in the following way.

To increase the number of alternatives, simply add alternatives $j^+$ with $p_{i,j^+}=0$ for all agents~$i$. 
These new alternatives do not affect the argument, as \eff implies range-respect, which ensures that none of them ever receives a positive amount. 
    
Adding agents is more involved, as our proof relies on explicit distributions induced by \prop and thus depends on the number of agents. However, we note that, throughout the proof, Agent $2$ always has the same peak, which puts all mass on alternative $b$.
Therefore, when adding agents~$i^+$ with $p_{i^+,b}=1$, we can run through the same proof but with adapted distributions. The generalized proof can be found in Appendix~\ref{app:impossforgeneraln}.
\end{proof}

\begin{remark}
\label{rem:range-respecting}
The bounds $m\ge 3$ and $n\ge 3$ in \Cref{thm:imposs} are tight. 
Indeed, there exists a moving phantom mechanism that satisfies strategyproofness, proportionality, and range-respect \citep[p.~22]{FPPV21a}, and it is known that range-respect and efficiency coincide when $m = 2$ or $n = 2$ \citep[Sec.~8]{EST23a}.

The three axioms required for the impossibility are independent.
Indeed, efficiency and strategyproofness (without proportionality) are satisfied by the mechanism that maximizes utilitarian welfare \citep{LNP08a};
strategyproofness and proportionality (without 
efficiency) are satisfied by the independent markets mechanism \citep{FPPV21a}; and proportionality and efficiency (without strategyproofness) are satisfied by 
a natural generalization of the maximum Nash welfare mechanism studied by \citet{ALLW25a}.
\end{remark}

\citet[p.~30]{FPPV21a} posed the question of whether every anonymous, neutral, continuous, and strategyproof mechanism can be represented as a moving phantom mechanism. 
While such a characterization might have the potential to simplify the previous proof, it does not hold in general; see Appendix~\ref{app:FPPVopenquestion}.

\subsection{$\ell_{\infty}$ preferences}

$\ell_1$ preferences take a special role among $\ell_p$ disutilities in terms of efficiency: indifference curves partially move along distributions with a constant sum on ``approved'' ($p_{i,j}>0$) alternatives. As an example, consider $M=\{a,b,c,d\}$ and an agent $i$ with peak $\p_i=(1/2,1/2,0,0)$. With $\ell_1$ preferences, she is indifferent between all distributions~$\q$ with $q_a+q_b=1/2$. This implies that if we have two agents and the second agent~$i'$ has $\p_{i'}=(0,0,0,1)$, every efficient distribution $\q$ with $q_a+q_b=1/2$ (equivalently, $q_c+q_d=1/2$) must put $0$ on alternative $c$ and $1/2$ on alternative $d$. By contrast, when considering, e.g., $\ell_2$ preferences, it also matters for agent $i$ \emph{how} $1/2$ is distributed on $c$ and $d$. As a result, more distributions become efficient, which weakens the role of efficiency for a potential impossibility when $p>1$.

We proceed by proving an impossibility for the preference model at the other end of the spectrum: $\ell_{\infty}$ preferences. 
These preferences behave similarly to $\ell_1$ disutilities (\Cref{obs:ell1}), which is helpful when arguing about efficiency.
\begin{observation}
\label{obs:ell8}
With $\ell_{\infty}$ preferences, if some agent $i$ has $p_{i,j}=1$, then for any distribution $\q$, $d_i(\q)=1-q_j$, regardless of the distribution on alternatives other than $j$. Therefore, agent $i$ is indifferent if some amount is moved between alternatives other than~$j$.
\end{observation}

\begin{theorem}\label{thm:imposslinf}
    With $\ell_{\infty}$ preferences, no mechanism satisfies efficiency, strategyproofness, and proportionality when $m \ge 3$ and $n \ge 3$.
\end{theorem}

\setcounter{table}{0}
The proof of \Cref{thm:imposslinf} can be found in Appendix~\ref{app:impossforgeneralnminf}.
It uses the same profiles as the one for \Cref{thm:imposs}, but needs more involved arguments when reasoning about efficiency and extending the argument to $m>3$  alternatives. 
The reason is that, with $\ell_{\infty}$ disutilities (in contrast to $\ell_1$),
efficiency does not imply range-respect,
and an efficient distribution might allocate a positive amount to an alternative to which all agents allocate $0$. 
For example, let $m=4$ and $n=2$ with peaks $(1/2,1/4,1/4,0)$ and $(1/4,1/2,1/4,0)$. Then, $(3/8,3/8,1/8,1/8)$ is efficient, as the maximal distance is $1/8$ for both agents, and if Agent $1$ is better off in distribution $\q$, this means $q_1>3/8$ which decreases Agent $2$'s utility.

\begin{remark}
\Cref{thm:imposslinf} requires $m\geq 3$, since for $m=2$, $\ell_1$ and $\ell_\infty$ preferences are single-peaked, and there are mechanisms that satisfy all requirements (see \Cref{rem:range-respecting}).
Moreover, $n\geq 3$ is required because, for $m=3$, the $\ell_\infty$ and $\ell_1$ metrics are equivalent---the $\ell_1$ distance is always twice the $\ell_\infty$ distance. 
Hence, the lower bound on $n$ has to be the same as in \Cref{thm:imposs}.

Similar to the impossibility for $\ell_1$ preferences, we expect all axioms to be independent.
However, to the best of our knowledge, this does not follow from existing results, as $\ell_\infty$ preferences have been studied significantly less than $\ell_1$.
\end{remark}

We conjecture that the incompatibility of efficiency, strategyproofness, and weak fairness conditions holds for $\ell_p$ disutilities for any $1 \le p \le \infty$, when $m\geq 3$ and $n\geq 3$.

In the next section, we demonstrate that the impossibility does not generalize to arbitrary peak-linear utility functions.

\section{The Nash product rule for Leontief preferences}
\label{sec:char}
In this section, we assume that preferences are representable by Leontief utilities. 
Inspired by the positive results obtained by maximizing the product of utilities in similar contexts \citep[in particular,][]{BGSS23a}, we define the \emph{Nash product rule} for budget aggregation as follows.
For any $P \in \mathcal{P}$,
\begin{align*}
    \nash[P]=\argmax_{\q \in \Delta^m}\prod_{i \in N}u_i(\q).
\end{align*}
\nash is well-defined as it always returns exactly one distribution \citep{BGSS23a}.

\subsection{Properties}

In this section, we investigate properties of \nash for budget aggregation.

Anonymity follows immediately from the fact that multiplication is commutative.
Neutrality is also straightforward as \nash does not take into account the identities of the alternatives.
Moreover, \nash satisfies continuity, which will be important for the axiomatic characterization we give in the next section. 

\begin{restatable}{proposition}{nashcont}
\label{prop:nashcontinuous}
    With Leontief utilities, \nash is continuous.
\end{restatable}

In addition, \nash is efficient and group-strategyproof \citep{BGSS23a}.\footnote{Our notion of group-strategyproofness is slightly different as agents do not have variable contributions. Specifically, we only need to consider manipulations of the individual peaks. However, this is a subset of all manipulations considered by \citet{BGSS23a}, so \nash also satisfies our notion of group-strategyproofness.}
Furthermore, \nash satisfies core fair share (and thus also proportionality), showing that the impossibilities for $\ell_1$ and $\ell_\infty$ preferences observed in \Cref{sec:imposs} do not apply for Leontief preferences.

\begin{restatable}{proposition}{nashcfs}
\label{prop:nash-cfs}
    With Leontief utilities, \nash satisfies core fair share.
\end{restatable}

The proofs of Propositions~\ref{prop:nashcontinuous} and \ref{prop:nash-cfs} can be found in Appendix \ref{subsec:nashprops}.

\subsection{Characterization}

In fact, \nash even admits an appealing characterization via strategyproofness and fairness. 

\begin{restatable}{theorem}{nashchar}
\label{thm:Nashchar}
With Leontief utilities,
\nash is the only continuous mechanism that satisfies group-strategyproofness and core fair share.
\end{restatable}

The proof of \Cref{thm:Nashchar} can be found in Appendix~\ref{subsec:nashcharproof}.
In light of the theorem, \nash appears to be the ideal mechanism for aggregating Leontief preferences.

As to the independence of the axioms, it is easy to see that core fair share is required for \Cref{thm:Nashchar} since any constant mechanism satisfies continuity and group-strategyproofness. The necessity of (group-)strategyproofness can be shown by slightly perturbing the outcome of \nash.
For example, consider $n=m=2$ and the two peaks at $1/4$ and $3/4$, respectively. \nash returns $\q=(1/2,1/2)$ with $u_i(\q)=2/3$. However, core fair share only guarantees utility $1/2$ for agent~$i$
(\Cref{lem:leontief-cfs}).
So any distribution which puts at least $3/8$ on both alternatives satisfies core fair share. This ``gap'' can be used to slightly change the outcome without violating core fair share.
Defining such changes in a continuous way (note that if both agents have the same peak, that change is 0) results in a different continuous mechanism satisfying core fair share.

We conjecture that continuity is required for the characterization as well.

The condition of group-strategyproofness is used only at one point of the proof (in \Cref{lem:keyprofsdistr}).
If we assume that agents have \emph{Leximin-Leontief preferences} (that is, subject to maximizing the smallest ratio, they maximize the second smallest ratio, etc.), then this condition can be weakened to ordinary strategyproofness; the proof is given in Appendix~\ref{sec:leximin}.

\begin{restatable}{theorem}{Nashcharlex}\label{thm:Nashcharlex}
With Leximin-Leontief preferences,
\nash is the only continuous mechanism that satisfies strategyproofness and core fair share.
\end{restatable}

\section{The case of two alternatives}\label{sec:m=2}

Notably, the impossibility results for $\ell_1$ and $\ell_\infty$ preferences require $m>2$. For $m=2$, the set of possible outcomes becomes one-dimensional and the class of all star-shaped utilities coincides with the well-studied class of \emph{single-peaked utilities}.

For two alternatives $M=\{a,b\}$, the set of outcomes can be identified with the unit interval $[0,1]$, where the endpoints $0$ and $1$ correspond to allocating the entire budget to alternatives $a$ and $b$, respectively. 

Denote by $\mathcal{U}^{SP}$ the set of all single-peaked utility functions.
We denote agent $i$'s peak~$p_i$ as a scalar in $[0,1]$ representing her favorite distribution $[1-p_i:a,\,p_i:b]$. 

In this section, we relax our assumption that $\mathcal{U}$ has to contain \emph{exactly} one utility function per peak in $[0,1]$; we only demand that $\mathcal{U}$ contains \emph{at least} one utility function per peak and implicitly assume that property from now on for all $\mathcal{U}\subseteq \mathcal{U}^{SP}$.
The ``at least one'' requirement is needed to allow agents to misreport their peak to any other peak in $[0,1]$. Note that we still require our mechanisms to be \emph{tops-only}, i.e., depend only on the agents' peaks. This generalization is possible because the mechanisms we characterize satisfy strategyproofness and further desirable properties without relying on any knowledge of the agents' utility functions except their peaks. 

Under the assumption that $\mathcal{U} = \mathcal{U}^{SP}$, \citet{Moul80a} characterized the set of all strategyproof mechanisms as generalized median rules. 
This characterization 
assumes that the rules have to handle \emph{all} profiles in $\mathcal{U}^{SP}$.
As a consequence, it no longer holds when restricting $\mathcal{U}$ to a strict subset of $\mathcal{U}^{SP}$, which often occurs in practical applications, e.g., when preferences are symmetric around the peak.
In principle, allowing rules to handle only a subset of $\mathcal{U}^{SP}$ may enable a greater selection of strategyproof rules.
This possibility has been studied in some later works \citep{BoJo83a,BeSe00a,MaMo11a,FPPV21a,ALLW25a}, which extended \citeauthor{Moul80a}'s result for some alternative axioms and to some specific subdomains of $\mathcal{U}^{SP}$.

We substantially generalize these results by proving that \citeauthor{Moul80a}'s characterization holds for \emph{any} subdomain 
$\mathcal{U}\subseteq \mathcal{U}^{SP}$,
when assuming continuity of the mechanism in addition.
The proofs are deferred to Appendix \ref{sec:proofsm=2}.

\begin{restatable}{theorem}{Moulsubdomain}
\label{thm:m=2}
For $m=2$ and arbitrary domain $\mathcal{U} \subseteq \mathcal{U}^{SP}$, 
a continuous mechanism $f$ satisfies anonymity and strategyproofness if and only if there exist $\alpha_0 \le \alpha_1 \le \dots \le \alpha_{n}$ in $[0,1]$ such that
    \begin{align*}
        f(P)=\med(p_1,\dots,p_n,\alpha_0, \dots, \alpha_{n}).
    \end{align*}
\end{restatable}

The \emph{uniform phantom rule} is a special case of the generalized median rule in which the peaks are distributed uniformly in $[0,1]$, that is, $\alpha_k=k/n$ for $k \in \{0,\dots,n\}$.
It is range-respecting, 
as at least $n+1$ arguments to the median 
(namely, $\alpha_0,p_1,\ldots,p_n$) 
are at most $\max_i p_i$,
and at least 
$n+1$ arguments to the median 
(namely, $\alpha_n,p_1,\ldots,p_n$) are at least $\min_i p_i$.

\citet{FPPV21a} showed that the uniform phantom rule is the only mechanism that ensures proportionality in addition to all axioms from \Cref{thm:m=2}.  
\citet{ALLW25a} strengthened this result by pointing out that continuity, strategyproofness, and proportionality suffice for characterizing the uniform phantom mechanism for symmetric single-peaked preferences.
Note that proportionality already contains some form of anonymity: when all agents have peaks at $[1:a]$ or $[1:b]$, proportionality requires picking a specific distribution that is independent of the agents' identities. 

Recently, \citet{JLPV24a} showed that the uniform phantom rule is the unique mechanism that satisfies strategyproofness and proportionality, among all rules defined on $\mathcal{U}^{SP}$. 
Again, we present a characterization that holds for every subset of $\mathcal{U}^{SP}$, whether symmetric or not.
\begin{restatable}{theorem}{unifphantomchar}
\label{thm:m=2+prop}
For $m=2$ and arbitrary domain $\mathcal{U}\subseteq \mathcal{U}^{SP}$, 
the only continuous mechanism that satisfies strategyproofness and proportionality is
 the uniform phantom rule.
\end{restatable}

\Cref{thm:m=2+prop} entails an interesting connection of \nash to the independent markets mechanism for $\ell_1$ preferences, which follows immediately from the mechanisms' axiomatic properties. 

\begin{corollary}
\label{prop:m=2_equiv_nash_l1}
With two alternatives, \nash for Leontief utilities is equivalent to the uniform phantom mechanism for $\ell_1$ preferences.  
\end{corollary}

\begin{proof}
For $m=2$, Leontief utilities as well as $\ell_1$ preferences are 
subsets of $\mathcal{U}^{SP}$, each of which contains one utility function per peak.
Both mechanisms are continuous, strategyproof, and proportional. Therefore, they need to be equivalent on their respective utility models by \Cref{thm:m=2+prop}.
\end{proof}

\section{Conclusion}

Aggregating individual distributions into a collective distribution constitutes an important problem in social choice theory.
Our work shows that understanding how agents form their preferences has crucial implications on the possibility of optimal aggregation mechanisms.

When agents' utilities are based on metrics such as $\ell_1$ and $\ell_\infty$, no rule simultaneously guarantees strategyproofness, efficiency, and proportionality.
However, when agents' utilities are non-metric and based on quotients (Leontief utilities), the Nash product rule guarantees group-strategyproofness and core fair share, which implies both efficiency and proportionality. Moreover, this rule is characterized by group-strategyproofness, core fair share, and continuity. 

The Nash product rule satisfies further desirable properties such as \emph{reinforcement} and \emph{participation}. The former states that when aggregating distributions for two disjoint electorates results in the same distribution, the mechanism should return the same distribution for the union of both electorates. The latter requires that agents are never better off by not participating in the aggregation mechanism.
Both statements follow trivially from the definition of the Nash product rule and hold for arbitrary utility models with nonnegative utilities.

It would be interesting to identify other sensible utility models for which the Nash product rule is a most attractive aggregation mechanism, and to pinpoint domain conditions that cause impossibilities similar to Theorems \ref{thm:imposs} and \ref{thm:imposslinf}.
Some concrete open questions are:
\begin{itemize}
    \item Does \Cref{thm:Nashchar} also hold when weakening group-strategyproofness to strategyproofness? Is continuity required for the characterization of the Nash product rule?
    \item Are there classes of peak-linear or star-shaped utility functions, other than Leontief, for which mechanisms satisfying core fair share and strategyproofness exist? In particular, for $\ell_p$ preferences with $1\leq p\leq \infty$ such as $\ell_2$, 
    are there strategyproof mechanisms that satisfy core fair share?
\end{itemize}

\subsection*{Acknowledgements}
This material is based on work supported by the Deutsche Forschungsgemeinschaft under grants BR 2312/11-2 and BR 2312/12-1, by the Israel Science Foundation under grant number 1092/24, by the Singapore Ministry of Education under grant number MOE-T2EP20221-0001, and by an NUS Start-up Grant. 
We thank anonymous referees as well as the participants of the 25th ACM Conference on Economics and Computation (July 2024), the Workshop on Voting, Matching, and Preference Aggregation at the National University of Singapore (December 2024), and the COMSOC Video Seminar (February 2025) for their insightful comments, stimulating discussions, and encouraging feedback.

\appendix

\section{Properties of Leontief utilities}
\label{sec:Leontiefcollection}

A useful concept when dealing with Leontief utilities is that of \emph{critical alternatives}.
\begin{definition}
\label{def:critical}
Given a distribution $\q$, 
we define the set of agent $i$'s \emph{critical alternatives}
\begin{align*}
T_{\q,i} \coloneqq 
\arg \min_{j\in M_i} \frac{q_j}{p_{i,j}}.
\end{align*}
\end{definition}
 
Critical alternatives allow for a characterization of efficient distributions.

\begin{lemma}[{\citet[Lem.~2]{BGSS23a}}]
\label{lem:efficient-iff-critical}
With Leontief utilities,
a distribution $\q$ is efficient if and only if every alternative $j$ with $q_j>0$ is critical for some agent. 
\end{lemma}

It is easy to see that core fair share implies a weak version of efficiency, which merely rules out the existence of distributions that make \emph{every} agent better off. For Leontief preferences, core fair share also implies the stronger version of efficiency used in this paper.

\begin{restatable}{proposition}{cfseff}
\label{prop:Leontief+cfs->eff}
    For Leontief utilities, core fair share implies efficiency.
\end{restatable}

\begin{proof}
Assume that $\q$ violates efficiency. So there is a distribution $\q'$ for which:
\begin{align}
\label{eq:cfs-violation}
  u_i(\q') \ge u_i(\q) && \text{for all $i \in N$, and}
 \\
 \notag
u_{i'}(\q') > u_{i'}(\q) && \text{for at least one $i' \in N$.}
\end{align}
The latter condition implies that every alternative $j\in T_{\q,i'}$ is allocated strictly more in $\q'$ than in $\q$. 

We now construct a new distribution $\q^*$ from $\q'$, by taking a small amount $m \epsilon$ from some $j_0\in T_{\q,i'}$, such that $q^*_{j_0} > q_{j_0}$ still holds, and then adding $\epsilon$ to every $j\in M$. Now we have
\begin{align}
\label{eq:q+>q}
    q^*_{j_0} &> q_{j_0},
    \\
    \notag
    q^*_{j} &> q'_{j} && \text{~for all } j\neq j_0.
\end{align}

We claim that $u_i(\q^*)>u_i(\q)$ for all $i\in N$.  Indeed, for all $i\in N$, if $j_0\in M_i$:
\begin{align*}
    u_i(\q^*) &= \min_{j\in M_i} ~ \frac{q^*_j}{p_{i,j}} && \text{(by definition of Leontief utilities)}
    \\
&= \min \left( \frac{q^*_{j_0}}{p_{i,j_0}}, ~ \min_{j\in M_i,\, j\neq j_0} ~ \frac{q^*_j}{p_{i,j}} \right) && \text{(by $\min$ properties)}
    \\        
    &> \min \left( \frac{q_{j_0}}{p_{i,j_0}}, ~ \min_{j\in M_i,\, j\neq j_0} ~ \frac{q'_j}{p_{i,j}} \right) && \text{(by \eqref{eq:q+>q})}
    \\
    &\geq \min \left( \frac{q_{j_0}}{p_{i,j_0}}, ~ \min_{j\in M_i} ~ \frac{q'_j}{p_{i,j}} \right) && \text{(by $\min$ properties)}
    \\
    &= \min \left( \frac{q_{j_0}}{p_{i,j_0}}, ~ u_i(\q')\right)  && \text{(by definition of Leontief utilities)}
    \\
    &\geq \min \left( \frac{q_{j_0}}{p_{i,j_0}}, ~ u_i(\q)\right)  && \text{(by \eqref{eq:cfs-violation})}
	\\
    &= 
    u_i(\q)  && \text{(by definition of Leontief utilities).}
\end{align*}
If $j_0\not\in M_i$, we can repeat a similar argument to arrive again at $u_i(\q^*) > u_i(\q)$.
Hence, the distribution $\q^*$ shows that $\q$ violates core fair share.
\end{proof}

\begin{remark}\label{rem:weakcfsLeon}
    A similar proof (where the set of all agents $N$ in (\ref{eq:cfs-violation}) is replaced by a subset $N'\subseteq N$ of a blocking coalition) shows that for Leontief utilities, core fair share is equivalent to the strong core fair share condition used by \citet{ABM20a}, where only one member of a blocking coalition has to become strictly better. 
\end{remark}

Another useful feature of Leontief utilities is that core fair share provides a uniform lower bound on all agents' utilities.
\begin{lemma}
\label{lem:leontief-cfs}
    With Leontief utilities, if $\q$ satisfies core fair share, then $u_i(\q)\geq 1/n$ for all $i\in[n]$.
\end{lemma}
\begin{proof}
If $u_i(\q)< 1/n$ for some $i\in[n]$, then $\{i\}$ is a blocking coalition by taking $\q' = \p_i$, as any distribution which gives at least $p_{i,j}/n$ for all $j\in[m]$ gives $i$ a utility of at least $1/n$.
\end{proof}

We now give a sufficient condition and another necessary condition for core fair share with Leontief utilities.
For any subset of agents $G\subseteq N$,
let $T_{\q,G}\coloneqq\bigcup_{i \in G}T_{\q,i}$ denote the set of alternatives critical to at least one agent from $G$.
\begin{lemma}
\label{lem:sufficient-for-cfs}
    With Leontief utilities, 
    if $\q(T_{\q,G}) \geq |G|/n$ for all subsets $G\subseteq N$,
    then $\q$ satisfies core fair share.
\end{lemma}
\begin{proof}
Assume for contradiction that $\q$ violates core fair share for some $G \subseteq N$. Then, there exists $\q' \in \Delta^m$ such that for every $\q'' \in \Delta^m$,
\begin{align*}
  u_i((|G|/n)\q'+(1-|G|/n)\q'') > u_i(\q) && \text{for all $i \in G$.}
 \end{align*}

Note that $T_{\q,G}=M$ cannot hold; otherwise, by \Cref{lem:efficient-iff-critical}, 
$\q$ would be efficient not only for $N$ but already for $G$, contradicting that $\q$ does not satisfy core fair share for $G$.
Therefore, there exists 
a distribution $\q''$ with $q''_j=0$
for every $j \in T_{\q,G}$.
Choosing such a distribution $\q''$ in the above inequality shows that $(|G|/n)q'_j > q_j$ for all $j\in T_{\q,G}$.
Thus, $\q(T_{\q,G})\coloneqq
\sum_{j \in T_{\q,G}}q_j 
< (|G|/n) \cdot \sum_{j \in T_{\q,G}}q'_j 
\le |G|/n
$. 
\end{proof}

The opposite direction of \Cref{lem:sufficient-for-cfs} does not hold even for $n=2$ and $m=2$. For example, suppose $\p_1 = (1/2,1/2)$, $\p_2 = (1/3,2/3)$, and $\q=(1/3,2/3)$. Then, $\q$ satisfies core fair share as the utility of each agent is at least $1/2$, but $\q(T_{\q,1}) = 1/3 < 1/2$.

\begin{lemma}
\label{lem:necessary-for-cfs}
With Leontief utilities, if 
$\q$ satisfies core fair share,
then 
$q_j=0$ if and only if $p_{i,j}=0$ for all $i \in N$.
\end{lemma}
\begin{proof}
If $q_j=0$ for some $j \in M$, then $u_i(\q)=0$ for all agents with $p_{i,j}>0$, meaning that core fair share is violated for each of these agents by \Cref{lem:leontief-cfs}. 

Conversely, $p_{i,j}=0$ for all $i \in N$ implies that $q_j=0$ for every efficient mechanism, where efficiency follows from \Cref{prop:Leontief+cfs->eff}. 
\end{proof}

We conclude this section by showing that core fair share is a stronger fairness axiom than proportionality for Leontief utilities, as well as for $\ell_1$ and $\ell_\infty$ preferences.
\begin{proposition}
\label{prop:cfs-implies-prop}
    With $\ell_1$ preferences,  $\ell_\infty$ preferences, or Leontief utilities, core fair share implies proportionality.
\end{proposition}

\begin{proof}
    Assume that a mechanism $f$ is not proportional for some single-minded profile $P \in \mathcal{P}$.
    Denote $\q \coloneqq f(P)$, and let  $N' \subseteq N$ be a maximal subset of agents where all agents in $N'$ allocate $1$ to the same alternative $j^*$ and proportionality is violated, i.e., $q_{j^*}< r$ for $r \coloneqq |N'|/n$. 

    Let $\q'$  be the peak of all agents $i\in N'$, i.e., $q'_{j^*} = 1$ and $q'_j = 0$ for all $j\ne j^*$. We claim that $u_i(r \q' + (1-r)\q'') > u_i(\q)$ for all distributions $\q''\in \Delta^m$.

    With $\ell_1$, $\ell_\infty$, and Leontief preferences, for all $i\in N'$,
    $u_i(\q)$ depends only on 
    $q_{j^*}$, and it is an increasing function of $q_{j^*}$.
    Specifically, 
    with $\ell_1$ preferences
    $u_i(\q)=-2(1-q_{j^*})$,
    with $\ell_\infty$ preferences
    $u_i(\q)=-(1-q_{j^*})$,
    and with Leontief preferences $u_i(\q) = q_{j^*}$.

Since $(r\q'+(1-r)\q'')_{j^*} \geq r > q_{j^*}$ for all $\q'' \in \Delta^m$, we have $u_i(r \q' + (1-r)\q'') > u_i(\q)$, so $f$ violates core fair share.
\end{proof}

As the proof shows, core fair share implies a property even stronger than proportionality: we do not need that $p_{i,j}\in \{0,1\}$ for all $i\in N$ but the guarantee is for every single-minded agent group, independently of the other agents' preferences.

\begin{remark}
The proof of \Cref{prop:cfs-implies-prop} does not work for $\ell_2$ preferences. For example, suppose $n=m=3$ and some two agents have their peak at $(1,0,0)$.
A rule that returns $\q = f(P)=(0.64,0.18,0.18)$ violates proportionality
but does not violate core fair share, as for both $i\in N'$, $u_i(\q)>-\sqrt{0.2}$,
but for $\q''=(0,1,0)$, we have $r\q'+(1-r)\q''=(2/3,1/3,0)$, which leads to $u_i = -\sqrt{2/9} < -\sqrt{0.2}$.
In fact, proportionality does not seem to be a very natural notion for such preferences, 
as the proportionality guarantee for single-minded agents, who put $1$ on alternative $j$, concerns only the distribution on alternative $j$,
whereas $\ell_2$ agents care also about the distribution on alternatives other than $j$.
\end{remark}

\section{Impossibility of efficiency, strategyproofness, and proportionality for $\ell_1$ preferences and arbitrary $n \ge 3$}
\label{app:impossforgeneraln}
\setcounter{table}{0}

In this section, we present the proof of \Cref{thm:imposs} for arbitrary $n \ge 3$ but still fixed $m=3$.  
\begin{proof}[Proof of \Cref{thm:imposs} continued]
Again, we set $M=\{a,b,c\}$ and write $\q=(q_a,q_b,q_c)$. 
Note that in contrast to the case $n=3$, we now also need to denote the number of agents with certain peaks in a profile.

Consider first the following two profiles.

        \begin{table}[h]
            \parbox{.49\linewidth}{
                \profileagent{1}
                  {$3/2n$ & $(2n-3)/2n$ & 0}{0 & 1 & 0}{0 & 0 & 1}
                  {{\ge} 1/2n}{{\ge} (2n-3)/2n}{{\le} 1/n}
                  \label{prof:1l1'}
            }
            \parbox{.49\linewidth}{
                \profileagent{2}
                  {1 & 0 & 0}{0 & 1 & 0}{0 & 0 & 1}
                  {1/n}{(n-2)/n}{1/n}
                  \label{prof:2l1'}
            }
            \end{table}

The outcome in Profile \ref{prof:2l1'} must be $(1/n,(n-2)/n,1/n)$ by \prop.
We now justify the bounds on the outcome in Profile \ref{prof:1l1'}. 
As Agent 1 can manipulate between Profile \ref{prof:1l1'} and Profile \ref{prof:2l1'}, \strp requires that Agent 1 does not gain from either manipulation. This implies that
\begin{align}
    d_1^{(1)}(\q^{(1)}) &\leq d_1^{(1)}(\q^{(2)}) = 2/n, \label{ineq:imposs1'}
    \\
    d_1^{(2)}(\q^{(1)}) &\geq d_1^{(2)}(\q^{(2)}) = (2n-2)/n. \label{ineq:imposs2'}
\end{align}

By (\ref{ineq:imposs1'}), $q^{(1)}_a \ge 1/2n$ (implying $q^{(1)}_b \le (2n-1)/2n$), $q^{(1)}_b \ge (2n-5)/2n$, and $q^{(1)}_c \le 1/n$.
By (\ref{ineq:imposs2'}), $q^{(1)}_a \le 1/n$, implying $q^{(1)}_b+q^{(1)}_c \ge (n-1)/n$, and thus $q^{(1)}_b \ge (n-2)/n$.

By \eff, we can even show that $q^{(1)}_b \ge (2n-3)/2n$. Otherwise, as $q^{(1)}_a > 0$, some small amount could be moved from $a$ to $b$. Agent $n$ is indifferent due to Observation~\ref{obs:ell1} and Agents $2,\dots,n-1$ strictly gain. Furthermore, this does not change agent $1$'s disutility as $q^{(1)}_b<(2n-3)/2n$.

        \begin{table}[h]
            \parbox{.49\linewidth}{
              \profileagentsmallmod{3}
                {$1/(n+1)$ & $n/(n+1)$ & 0}{0 & 1 & 0}{0 & 0 & 1}
                {0}{(n-1)/n}{1/n}
                \label{prof:3l1'}
            }
            \parbox{.49\linewidth}{
              \profileagentsmall{4}
                {0 & 1 & 0}{0 & 1 & 0}{0 & 0 & 1}
                {0}{(n-1)/n}{1/n}
                \label{prof:4l1'}
            }
            \end{table}
            
Next, we consider Profiles 3 and 4.
The outcome in Profile \ref{prof:4l1'} follows from \prop.
We now prove that the outcome in Profile \ref{prof:3l1'} must be the same. 
As Agent $1$ can manipulate between Profile \ref{prof:3l1'} and Profile \ref{prof:4l1'}, \strp requires that Agent 1 does not gain from either manipulation. This implies that
\begin{align}
    d_1^{(3)}(\q^{(3)}) &\leq d_1^{(3)}(\q^{(4)}) = 2/n, \label{ineq:imposs3'}
    \\
    d_1^{(4)}(\q^{(3)}) &\geq d_1^{(4)}(\q^{(4)}) = 2/n. \label{ineq:imposs4'}
\end{align}

By (\ref{ineq:imposs3'}), $q^{(3)}_c \le 1/n$, implying $q^{(3)}_a+q^{(3)}_b \ge (n-1)/n$.
By (\ref{ineq:imposs4'}), $q^{(3)}_b \le (n-1)/n$.

However, by \eff, if $q^{(3)}_a>0$ then $q^{(3)}_b\geq n/(n+1)$. 
Otherwise, some small amount can be moved from $a$ to $b$. 
Agent $n$ is indifferent due to Observation \ref{obs:ell1} and Agents $2,\dots,n-1$ strictly gain. Furthermore, this does not change Agent $1$'s disutility as $q^{(3)}_b<n/(n+1)$.
Therefore, $q^{(3)}_a=0$ must hold, and the only outcome compatible with \strp is $\q^{(3)} = (0, (n-1)/n, 1/n)$.

Now that we know $\q^{(3)}$, we 
consider a manipulation of Agent $1$ from Profile \ref{prof:3l1'} to Profile~\ref{prof:1l1'}.
Strategyproofness implies 
\begin{align*}
   d_1^{(3)}(\q^{(1)}) \geq d_1^{(3)}(\q^{(3)}) = 2/n. 
\end{align*}
But the bounds we already have for $\q^{(1)}$ 
imply that $d_1^{(3)}(\q^{(1)}) \leq 2/n$ as $q^{(1)}_a \ge 1/2n$ and $q^{(1)}_b \ge (2n-3)/2n$. Therefore, $d_1^{(3)}(\q^{(1)}) = 2/n$ together with $q^{(1)}_a = 1/2n$ and $q^{(1)}_b=(2n-3)/2n$.
Hence, $\q^{(1)} = (1/2n, (2n-3)/2n, 1/n)$.    

Finally, we consider the following two profiles.

        \begin{table}[h]
            \parbox{.49\linewidth}{
                \profileagentsmall{5}
                  {$3/2n$ & $(2n-3)/2n$ & 0}{0 & 1 & 0}{0 & $(2n-3)/2n$ & $3/2n$}
                  {}{}{}
                  \label{prof:5l1'}
                }
            \parbox{.49\linewidth}{
                \profileagentsmall{6}
                  {1 & 0 & 0}{0 & 1 & 0}{0 & $(2n-3)/2n$ & $3/2n$}
                  {1/n}{(2n-3)/2n}{1/2n}
                  \label{prof:6l1'}
                }
            \end{table}

The distribution $\q^{(6)}$ is determined by arguments analogous to those for $\q^{(1)}$, reasoning about Agent $n$ instead of Agent 1.

We now consider a manipulation of Agent 1 from Profile \ref{prof:5l1'} to Profile \ref{prof:6l1'}.
It follows from \strp that
\begin{align*}
    d_1^{(5)}(\q^{(5)})\leq d_1^{(5)}(\q^{(6)})=1/n,
\end{align*}
which implies that $q^{(5)}_c \le 1/2n$.
Similarly, we consider a manipulation of Agent $n$ from Profile \ref{prof:5l1} to Profile \ref{prof:1l1}. 
It follows from \strp that
\begin{align*}
    d_n^{(5)}(\q^{(5)})\leq d_n^{(5)}(\q^{(1)})=1/n,
\end{align*}
which implies that $q^{(5)}_c \ge 3/2n-1/2n=1/n$, a contradiction.
\end{proof}

\section{Not every anonymous, neutral, continuous, and strategyproof rule is a moving phantom mechanism}
\label{app:FPPVopenquestion}

For a formal definition of moving phantom mechanisms, see Definition 6 of \citet{FPPV21a}.

\begin{proposition}
\label{prop:moving-phantom}
    With $\ell_1$ preferences, not every anonymous, neutral, continuous, and strategyproof mechanism can be represented as a moving phantom mechanism, for any $n\geq 1$ and $m\geq 3$ and for any number of phantom functions.
\end{proposition}

\begin{proof}
We first prove the claim for $n=1$ and $m=3$.
Consider the mechanism which, in general, returns the agent's peak but cannot put more than $0.9$ on an alternative. If (without loss of generality) $p_{1,1}>0.9$, the mechanism returns $q_1=0.9$, $q_2=p_{1,2}+(p_{1,1}-0.9)/2$, and $q_3=p_{1,3}+(p_{1,1}-0.9)/2$.
Since this outcome minimizes the $\ell_1$ distance of Agent $1$ among all ``legal'' distributions, the mechanism is strategyproof. 

Anonymity is trivially satisfied as there is only one agent.
For neutrality, if $p_{1,j}=p_{1,k}$, then both alternatives receive the same probability share. In particular, if $p_{1,l}>0.9$ for the third alternative $l$, then $j$ and $k$ both receive $p_{1,j}+(p_{1,l}-0.9)/2$, and the distribution of the surplus does not depend on the identity of the alternatives. 
For continuity, the only ``critical'' points are those where $p_{1,j}> 0.9$ approaches $0.9$ from above. For such peaks, $q_j=0.9$ is constant and the ``surplus'' $p_{1,j}-0.9$ is distributed on the other two alternatives in a continuous manner. Thus, the mechanism also satisfies continuity.

Suppose by contradiction that the above mechanism can be represented as a moving phantom mechanism with phantom functions $\textbf{h}$; let $k$ be the number of phantoms.
Let $\p_1=(0.91,0.08,0.01)$. Given this profile, the mechanism returns $\q=(0.9,0.085,0.015)$.
This implies that, for some $t\in[0,1]$, 
$0.085$ is the median of $0.08$ and $h_1(t),\ldots,h_{k}(t)$,
so the number of phantoms larger than or equal to $0.085$ should be at least $k/2+1$ (for even $k$) or $(k+3)/2$ (for odd $k$).%
\footnote{
We assume that when $k$ is odd, the median of $k+1$ elements is the ($(k+1)/2$)-th element.
}
By similar considerations, since
$0.015$ is the median of $0.01$ and $h_1(t),\ldots,h_{k}(t)$,
the number of phantoms smaller than or equal to $0.015$ should be at least $k/2$ (for even $k$)
or $(k-1)/2$ (for odd $k$).
These two observations are contradictory as there are only $k$ phantoms in total.

A similar construction also works for larger $m$. Moreover, since the considered properties do not relate instances with different $n$, such a construction can be extended to a rule for arbitrary $n$ by using this mechanism when $n=1$ and a moving phantom mechanism when $n \ge 2$.
\end{proof}
Note that the proof of \Cref{prop:moving-phantom} does not assume continuity or any other property of the phantom functions. 

After constructing this counterexample, we learned that \citet{BFS+24a} independently came up with a similar construction with a more natural extension to larger $n$, that does not coincide with a moving phantom mechanism for $n>1$.

\section{Impossibility of efficiency, strategyproofness and proportionality for $\ell_\infty$ preferences and $n,m \ge 3$}
\label{app:impossforgeneralnminf}
\setcounter{table}{0}

In this section, we present the proof of \Cref{thm:imposslinf} for arbitrary $n \ge 3$ and $m \ge 3$. We start by fixing $m=3$ and considering $n \ge 3$.

\begin{lemma}
\label{lem:impossinfm=3}
    With $\ell_{\infty}$ preferences, no mechanism satisfies efficiency, strategyproofness, and proportionality when $m = 3$ and $n \ge 3$.
\end{lemma}

\begin{proof}
We use the same notation as in the proof of \Cref{thm:imposs}. 

        \begin{table}[h]
            \parbox{.49\linewidth}{
                \profileagent{1}
                  {$3/2n$ & $(2n-3)/2n$ & 0}{0 & 1 & 0}{0 & 0 & 1}
                  {{\ge} 1/2n}{{\ge} (2n-3)/2n}{{\le} 1/n}
                  \label{prof:1linf}
            }
            \parbox{.49\linewidth}{
                \profileagent{2}
                  {1 & 0 & 0}{0 & 1 & 0}{0 & 0 & 1}
                  {1/n}{(n-2)/n}{1/n}
                  \label{prof:2linf}
            }
            \end{table}

Consider first Profiles 1 and 2.
The outcome in Profile \ref{prof:2linf} must be $(1/n,(n-2)/n,1/n)$ by \prop.
We now justify the bounds on the outcome in Profile \ref{prof:1linf}. 
As Agent 1 can manipulate between Profile~\ref{prof:1linf} and Profile \ref{prof:2linf}, \strp requires that Agent 1 does not gain from either manipulation. This implies that
\begin{align}
    d_1^{(1)}(\q^{(1)}) &\leq d_1^{(1)}(\q^{(2)}) = 1/n, \label{ineq:imposs1inf}
    \\
    d_1^{(2)}(\q^{(1)}) &\geq d_1^{(2)}(\q^{(2)}) = (n-1)/n. \label{ineq:imposs2inf}
\end{align}

By (\ref{ineq:imposs1inf}), $q^{(1)}_a \ge 1/2n$ (implying $q^{(1)}_b \le (2n-1)/2n$), $q^{(1)}_b \ge (2n-5)/2n$, and $q^{(1)}_c \le 1/n$.
By (\ref{ineq:imposs2inf}), $q^{(1)}_a \le 1/n$, implying $q^{(1)}_b+q^{(1)}_c \ge (n-1)/n$, and thus $q^{(1)}_b \ge (n-2)/n$.

By \eff, we can even show that $q^{(1)}_b \ge (2n-3)/2n$. 
Otherwise, $q^{(1)}_b < (2n-3)/2n$ and $q^{(1)}_c + q^{(1)}_a > 3/2n$, and some small amount can be moved from $a$ to $b$.
Agent $n$ is indifferent due to Observation~\ref{obs:ell8} and Agents $2,\dots,n-1$ strictly gain.
Furthermore, this does not increase Agent $1$'s disutility as $d_1^{(1)}(\q^{(1)}) \ge q^{(1)}_c > 3/2n-q^{(1)}_a$ and $q^{(1)}_b < (2n-3)/2n$.
Hence, $q^{(1)}_b \ge (2n-3)/2n$. 

        \begin{table}[h]
            \parbox{.49\linewidth}{
              \profileagent{3}
                {$1/(n+1)$ & $n/(n+1)$ & 0}{0 & 1 & 0}{0 & 0 & 1}
                {}{}{}
                \label{prof:3linf}
            }
            \parbox{.49\linewidth}{
              \profileagentsmall{4}
                {0 & 1 & 0}{0 & 1 & 0}{0 & 0 & 1}
                {0}{(n-1)/n}{1/n}
                \label{prof:4linf}
            }
            \end{table}

Assume for contradiction that $q^{(1)}_c \le 3/4n$.

Consider a manipulation of Agent $1$ from Profile \ref{prof:3linf} to Profile \ref{prof:1linf}. Note that $d^{(3)}_1(\q^{(1)})\le 3/4n$ with the bounds established for $\q^{(1)}$. 
By \strp for Agent $1$, $q^{(3)}_c \le 3/4n$. 

By \eff, $q^{(3)}_b \ge n/(n+1)$.
Otherwise, $q^{(3)}_b < n/(n+1)$ and $q^{(3)}_a > 1/(n+1)-q^{(3)}_c$, and some small amount can be moved from $a$ to $b$.
Agent $n$ is indifferent due to Observation \ref{obs:ell8} and Agents $2,\dots,n-1$ strictly gain. Furthermore, this does not increase Agent $1$'s disutility as $d_1^{(3)}(\q^{(3)}) \ge q^{(3)}_c > 1/(n+1)-q^{(3)}_a$ and $q^{(3)}_b < n/(n+1)$.
Hence, $q^{(3)}_b \ge n/(n+1)$.
However, as $n/(n+1)>(n-1)/n$, this contradicts \strp for Agent $1$ manipulating from Profile \ref{prof:4linf} to Profile \ref{prof:3linf}, where $\q^{(4)}=(0,(n-1)/n,1/n)$ follows from \prop.

        \begin{table}[h]
            \parbox{.49\linewidth}{
                \profileagentmod{5}
                  {1 & 0 & 0}{0 & 1 & 0}{0 & $(2n-3)/2n$ & $3/2n$}
                  {{\le} 1/n}{{\ge} (2n-3)/2n}{{\ge} 1/2n}
                  \label{prof:5linf}
                }
            \parbox{.49\linewidth}{
                \profileagentsmall{6}
                  {$3/2n$ & $(2n-3)/2n$ & 0}{0 & 1 & 0}{0 & $(2n-3)/2n$ & $3/2n$}
                  {}{}{}
                  \label{prof:6linf}
                }
            \end{table}

Therefore, $q^{(1)}_c>3/4n$ has to hold. By analogous arguments with reversed roles of Agents $1$ and~$n$, the same bounds for $\q^{(5)}$ as well as $q^{(5)}_a>3/4n$ hold. 

Since $q^{(1)}_b \ge (2n-3)/2n$, we must have $q^{(1)}_a < 3/4n$. 
Consider a manipulation of Agent~$n$ from Profile \ref{prof:6linf} to Profile \ref{prof:1linf}. Note that $d^{(6)}_n(\q^{(1)})<3/4n$, as $q^{(1)}_b \ge (2n-3)/2n$ and $q^{(1)}_c>3/4n$. By \strp for agent $n$, we have $d^{(6)}_n(\q^{(6)}) \le d^{(6)}_n(\q^{(1)})<3/4n$, and thus $q^{(6)}_a<3/4n$.
Finally, consider a manipulation of Agent $1$ from Profile \ref{prof:6linf} to Profile~\ref{prof:5linf}. Analogously, $d^{(6)}_1(\q^{(6)}) \le d^{(6)}_1(\q^{(5)})<3/4n$, which implies $q^{(6)}_a > 3/2n-3/4n=3/4n$, a contradiction.  
\end{proof}

We finish the proof of \Cref{thm:imposslinf} by showing that this argument remains valid under the addition of alternatives $j^+$ with $p_{i,j^+}=0$ for all agents $i \in N$.

\begin{proof}[Proof of \Cref{thm:imposslinf}]
    We prove that no efficient mechanism puts positive probability on new alternatives $j^+$ with $p_{i,j^+}=0$ for all agents $i \in N$ in any of the six profiles used for the proof of \Cref{lem:impossinfm=3}. Together with \Cref{lem:impossinfm=3}, this completes the proof of \Cref{thm:imposslinf}.

    Proportionality directly implies that adding such an alternative $j^+$ to Profiles \ref{prof:2linf} and \ref{prof:4linf} does not change the distribution.

    Next, consider Profile \ref{prof:1linf}. 
    If $|(2n-3)/2n-q^{(1)}_{b}|<d_1^{(1)}(\q^{(1)})$, or $|(2n-3)/2n-q^{(1)}_{b}|=d_1^{(1)}(\q^{(1)})$ and $q^{(1)}_b<(2n-3)/2n$, moving some amount of probability from $j^+$ to $b$ cannot increase Agent~$1$'s disutility, does not change Agent $n$'s disutility by \Cref{obs:ell8}, and decreases the disutilities of all other agents. Therefore, such a redistribution would correspond to a Pareto improvement.
    Note that $|(2n-3)/2n-q^{(1)}_{b}|=d_1^{(1)}(\q^{(1)})$ and $q^{(1)}_b=(2n-3)/2n$ cannot hold simultaneously, as $q^{(1)}_{j^+}>0$.
    Hence, the only remaining case we need to consider is $|(2n-3)/2n-q^{(1)}_{b}|=d_1^{(1)}(\q^{(1)})$ and $q^{(1)}_b>(2n-3)/2n$.
    This implies $3/2n-q^{(1)}_a \le d_1^{(1)}(\q^{(1)})$ and thus, $q^{(1)}_a+q^{(1)}_b \ge 3/2n-d_1^{(1)}(\q^{(1)})+(2n-3)/2n+d_1^{(1)}(\q^{(1)})=1$, contradicting $q^{(1)}_{j^+}>0$.
    The argument for Profile \ref{prof:5linf} works analogously.

    For Profile \ref{prof:3linf}, moving some amount of probability from $j^+$ to $b$ can only potentially increase the disutility of one agent, namely Agent $1$, if $q^{(3)}_b \ge n/(n+1)$ and $d_1^{(3)}(\q^{(3)})=q^{(3)}_b-n/(n+1)$. But then, $1/(n+1)-q^{(3)}_a \le d_1^{(3)}(\q^{(3)})$ and again, $q^{(3)}_a+q^{(3)}_b=1$, contradicting $q^{(3)}_{j^+}>0$.

    Finally, for Profile \ref{prof:6linf}, moving some amount of probability from $j^+$ to $b$ can only potentially increase the disutilities of two agents, namely Agents $1$ and $n$, if $q^{(6)}_b \ge (2n-3)/2n$ and $d_k^{(6)}(\q^{(6)})=q^{(6)}_b-(2n-3)/2n$ holds for at least one $k \in \{1,n\}$, without loss of generality for $k=1$. But then, $3/2n-q^{(6)}_a \le d_1^{(6)}(\q^{(6)})$ and again, $q^{(6)}_a+q^{(6)}_b=1$, contradicting $q^{(6)}_{j^+}>0$. 
\end{proof}

\section{Proofs for \Cref{sec:char}}\label{sec:nashproofs}

Analogously to its application in donor coordination \citep[][]{BGSS23a}, we can also interpret the outcome of \nash as a Nash equilibrium where each agent $i$ reports a vector $\s_i \in \Delta^m$ and the outcome is determined by adding up the score $\sum_{i \in N}s_{i,j}$ of each alternative $j$. Here, the strategy set is the set of preferences. 
This interpretation will be useful for proving certain properties of \nash. 

\begin{definition}[Decomposition]
A \emph{decomposition} of a distribution $\q$ is a vector of nonnegative score vectors $(\s_i)_{i \in N}$ with
\begin{align*}
&\sum_{i \in N}s_{i,j} = q_j && \text{ for all } j\in M;
\\
&\sum_{j \in M}s_{i,j} = \frac{1}{n} && \text{ for all } i\in N.
\end{align*}
\end{definition}

An alternative characterization of the \nash outcome uses the notion of critical alternatives. 
Recall the definition of this notion from Appendix~\ref{sec:Leontiefcollection}.

\begin{lemma}[{\citet[Sec.~4.3]{BGSS23a}}]
\label{lem:eq-iff-critical}
A distribution $\q$ maximizes the Nash product
if and only if
it has a decomposition $(\s_i)_{i \in N}$ such that $s_{i,j}=0$ for every alternative $j\not\in T_{\q,i}$.
\end{lemma}

Another fact to keep in mind is that utilities and efficient outcomes admit a one-to-one correspondence.

\begin{lemma}[{\citet[Lem.~3]{BGSS23a}}]
\label{lem:efficientunique}
Let $\q$ and $\q'$ be efficient distributions inducing the same utility vector, that is, $u_i(\q)=u_i(\q')$ for all $i \in N$. Then, $\q=\q'$.
\end{lemma}

With $\ell_1$ preferences, every efficient distribution $\q$ must be range-respecting (that is, $q_j$ must be between $\min_i p_{i,j}$ and $\max_i p_{i,j}$ for all $j\in M$).
However, with Leontief utilities, this is not the case. 
In fact, \nash is efficient but not range-respecting. 
Intuitively, \nash may prefer to decrease the distribution for one alternative below the minimum peak, in order to increase the distribution for other alternatives whose ratio is smaller.
Nevertheless, an efficient distribution always satisfies one direction of range-respect.
\begin{lemma}
\label{lem:one-sided-range-respect}
    With Leontief utilities, if $\q$ is an efficient distribution, then $q_j \leq \max_i p_{i,j}$ for all $j\in M$.
    In particular, this holds for any distribution returned by \nash.
\end{lemma}
\begin{proof}
    Suppose by contradiction that, for some $j\in M$,  
    it holds that $q_j > p_{i,j}$ for all $i\in N$.
    Construct a new distribution $\q'$ from $\q$ by removing some amount from $j$ and allocating it equally among the other alternatives, such that $q_j > p_{i,j}$ for all $i\in N$ still holds.

    Note that, for any agent $i$ with Leontief utilities, $u_i(\q)\leq 1$ for all $\q$. Therefore, the decrease in $q_j$ does not decrease the Leontief utility of any agent, as $q_j/p_{i,j} > 1$. But the increase in allocation to other alternatives must increase the utilities of all agents. Therefore, $\q'$ is a Pareto improvement of $\q$, contradicting efficiency.    
\end{proof}

\subsection{Properties of \nash}\label{subsec:nashprops}

\nashcont*

\begin{proof}
Suppose we are given a sequence of profiles $P^1$, $P^2$, $\dots$ converging to $P^*$, i.e., $\lim_{k \to \infty} p^k_{i,j}=p^*_{i,j}$ for every agent $i \in N$ and alternative $j \in M$. 
Denote by $u^k_i$ the utility of agent~$i$ in profile $P^k$, and $u^*_i$ the utility in profile $P^*$.
Denote $\q^k=\nash[P^k]$ for every $k \in \mathbb{N}$ and $\q^*=\nash[P^*]$. By boundedness, it suffices to show that every convergent subsequence of $\q^1$, $\q^2$, $\dots$ converges to $\q^*$. Take such a subsequence, which must exist by the Bolzano-Weierstrass theorem, and denote its limit by $\q^L$. With abuse of notation, we now refer to this subsequence as $\q^1$, $\q^2$, $\dots$. Our goal is to show that $\q^L = \q^*$.

\emph{Case 1:
$q^*_j>0$ for all alternatives $j \in M$.}
Denote by $\nash[\q,P]$ the Nash welfare of the outcome~$\q$ when the profile is $P$. 
By definition of \nash on $P^k$, we have 
$$\nash[\q^*, P^k] \le \nash[\q^k, P^k]$$ for every $k$.
We take the limit of both sides as $k\to\infty$.
\begin{itemize}
\item
The left-hand side is a product of utilities $u_i^k(\q^*)$, where the distribution is fixed and only the utility functions change. Each utility is a minimum of ratios $q^*_j/p^k_{i,j}$ where all numerators are at least $\epsilon$, for some $\epsilon>0$. 
Since the minimum is always at most $1$, elements with $p^k_{i,j}<\epsilon$ do not affect the minimum and can be ignored. Therefore, the minimum is determined only by ratios with $p^k_{i,j}\geq \epsilon$. In this domain, the ratios are continuous functions of $p_{i,j}^k$, and their minimum is continuous too. Therefore, $\lim_{k\to\infty} u_i^k(\q^*) = u_i^*(\q^*)$,
so the limit of the product at the left-hand side equals $\nash[\q^*, P^*]$. 
\item 
The right-hand side is a product of utilities $u_i^k(\q^k)$, where both the distribution and the utility functions change.
There may be agents $i$ and alternatives $j$ for which both $q^k_j$ and $p^k_{i,j}$ approach~$0$ as $k \to \infty$, so the limit of $u_i^k(\q^k)$ may differ from $u_i^*(\q^L)$.
For example, if $\p_i^k = (2/k,1-2/k)$ and $\q^k = (1/k,1-1/k)$ then $u_i^k(\q^k) = 1/2$ for all $k$, but $\q^L=(0,1)$ so $u_i^*(\q^L) = 1$.
However, any alternative $j$ for which 
$p^*_{i,j}=\lim_{k \to \infty}p^k_{i,j}=0$ 
is removed from the minimum, 
so the minimum at the limit profile can only be larger than the limit of minima.
Therefore, $\lim_{k\to\infty} u_i^k(\q^k) \leq u_i^*(\q^L)$,
and the limit of the product at the right-hand side is at most $\nash[\q^L, P^*]$.

\end{itemize}

Therefore, we have $\nash[\q^*, P^*] \le \nash[\q^L, P^*]$. By definition and uniqueness of \nash on~$P^*$, we get $\q^L=\q^*$.

\emph{Case 2: $q^*_j=0$ for some alternatives $j\in M$.} Define $Z=\{j \in M:q^*_j=0\}$. As $\q^*$ maximizes Nash welfare for $P^*$, we have $p^*_{i,j}=0$ for every $ i \in N$ and $j \in Z$; otherwise, an agent $i$ with $p^*_{i,j}>0$ would receive zero utility causing the whole product to become zero.
Consequently, $\lim_{k\to\infty} p^k_{i,j} = 0$ for every $i \in N$, $j \in Z$.
By \Cref{lem:one-sided-range-respect}, the amount allocated to alternatives in $Z$ by $\q^k$ also tends to zero for $k \to \infty$. 
Hence, $q^*_j=0$ implies $q^L_j=0$.

Now, we will measure utilities and Nash welfare only with respect to the alternatives outside $Z$. Let $w^k$ be the total amount allocated to alternatives outside $Z$ in $\q^k$, so we know that $w^k$ converges to $1$ for $k \to \infty$. By definition of \nash on $P^k$, we have 
$$\nash[w^k \cdot \q^*, P^k] \le \nash[\q^k, P^k].$$ 

As in Case 1, we take the limit of both sides, the left-hand side equals $\nash[\q^*, P^*]$, and the right-hand side is at most $\nash[\q^L, P^*]$. Since in $P^*$ all agents assign zero to alternatives in $Z$, these two quantities remain the same even if we take the alternatives in~$Z$ back into account for the Nash welfare. Hence, as in Case 1, we get $\nash[\q^*, P^*] \le \nash[\q^L, P^*]$. By definition and uniqueness of \nash on $P^*$, we conclude that $\q^L = \q^*$. 
\end{proof}

\nashcfs*
\begin{proof}
Let $G$ be any subset of agents.
By \Cref{lem:eq-iff-critical}, 
the \nash distribution can be decomposed in such a way that
every agent from $G$ only contributes her share of $1/n$ to alternatives in $T_{\q,G}$. Thus, $\q(T_{\q,G}) \ge |G|/n$. 
By \Cref{lem:sufficient-for-cfs}, \nash satisfies core fair share.
\end{proof}

\subsection{Characterization of \nash}
\label{subsec:nashcharproof}

\nashchar*

Let $f$ be a mechanism satisfying the properties in the theorem statement.
The proof is divided into three lemmas and has the following structure. Starting at an arbitrary profile $P$, we first show in \Cref{lem:keyprofsdistr} that moving to a ``key'' profile $P^*$ cannot change the outcome: $f(P^*)=f(P)$. Then, \Cref{lem:keyprofs} states that $f(P^*)=\nash[P^*]$. Finally, \Cref{lem:keyprofseq} proves $\nash[P^*]=\nash[P]$, which completes the proof as we then have $f(P)=\nash[P]$. 

Let $\q\coloneqq f(P)$. By core fair share
and \Cref{lem:necessary-for-cfs},  $q_j=0$ if and only if $p_{i,j}=0$ for all $i \in N$. 

Denote by $P^*$ the profile with peaks
    \begin{align*}
        p^*_{i,j}=
        \begin{cases}
            p_{i,j} /\p_i(T_{\q,i}) &\text{for $j \in T_{\q, i}$};\\
            0 &\text{for $j \not \in T_{\q, i}$},
        \end{cases}
    \end{align*}
    where $\displaystyle \p_i(T_{\q,i})\coloneqq\sum_{j \in T_{\q,i}}p_{i,j}$.
That is, in $P^*$, each agent moves her peak so that it is nonzero only on alternatives critical for her under $\q$.
For example, suppose $\p_i = (0.1, 0.2, 0.3, 0.4)$ and $\q = (0.1, 0.1, 0.6, 0.2)$. Then $T_{\q,i}=\{2,4\}$, $\p_i(T_{\q,i})=0.6$, and $\p^*_i = (0, 1/3, 0, 2/3)$.
Note that, for an agent $i$ with $T_{\q,i}=M$, it holds that $\p_i=\p^*_i$. 

\begin{lemma}\label{lem:keyprofsdistr}
With Leontief utilities, if $f$ is a continuous mechanism satisfying group-strategyproofness and efficiency,
then

(a) the outcome does not change, that is, $f(P^*)=f(P)=\q$;

(b) the sets of critical alternatives do not change, that is, $T_{\q,i}=T^*_{\q,i}$ for every $i \in N$.
\end{lemma}

\newcommand{\comb}[1]{\widehat{#1}}

\begin{proof}
(a) 
We move the peak of each agent in turn.
For each agent $i$, we change $\p_i$ towards $\p^*_i$ gradually, to some  $\comb{\p}_i \coloneqq \lambda \p_i^*+(1-\lambda)\p_i$, for some $\lambda \in [0,1]$ to be computed later. 
Then we proceed along this line until we reach $\p^*_i$.
In the above example, $\lambda=0.3$ gives 
$\comb{\p}_i = (0.07, 0.24, 0.21, 0.48)$.
If $\p_i = \p^*_i$, it is clear that the outcome does not change, so assume that $\p_i \ne \p^*_i$.
The change from $\p_i$ to $\comb{\p}_i$ has a simple structure:
\begin{itemize}
\item 
$\comb{p}_{i,j} > p_{i,j}$ for all $j\in T_{\q,i}$,
and the ratio $\comb{p}_{i,j} / p_{i,j} = \lambda/\p_i(T_{\q,i}) + (1-\lambda) =: \lambda^+$, a constant independent of $j$ (in the example, $\lambda^+=1.2$);
\item 
$\comb{p}_{i,j} < p_{i,j}$ for all $j\not \in T_{\q,i}$, and the ratio 
$\comb{p}_{i,j} / p_{i,j} = (1-\lambda) =: \lambda^-$, again independent of $j$ (in the example, $\lambda^-=0.7$).
\end{itemize}
Now, consider the ratios $q_j/p_{i,j}$ versus the ratios $q_j/\comb{p}_{i,j}$.
For each  $j\in T_{\q,i}$, we have
$q_j/p_{i,j} > q_j/\comb{p}_{i,j}$ because $\comb{p}_{i,j} > p_{i,j}$, whereas for each $j\not\in T_{\q,i}$, we have $q_j/p_{i,j} < q_j/\comb{p}_{i,j}$ because $\comb{p}_{i,j} < p_{i,j}$.
Furthermore, for all $j\in T_{\q,i}$, the ratios $q_j/\comb{p}_{i,j}$ remain equal (as $\comb{p}_{i,j} / p_{i,j}$ is constant)  when moving from $\p_i$ to $\comb{\p}_i$,
and they remain the smallest ratios for agent $i$. This implies that $\comb{T}_{\q,i} = T_{\q,i}$.

Moreover, the entire ordering of alternatives 
by the ratio $q_j/p_{i,j}$
is identical to the ordering of alternatives 
by the ratio $q_j/\comb{p}_{i,j}$,
as the smallest ratio is divided by $\lambda^+>1$ and the other ratios are divided by $\lambda^-<1$.
In other words, suppose we partition the alternatives into subsets according to the ratio $q_j/p_{i,j}$, 
and denote the subset with the smallest ratio by $T_{\q,i,1} \equiv T_{\q,i}$, the subset with the second-smallest ratio by $T_{\q,i,2}$, etc., then 
$T_{\q,i,r} = \comb{T}_{\q,i,r}$ for all $r\geq 1$.

\emph{Computing $\lambda$.}
We pick $\lambda$ sufficiently small such that no new alternative becomes critical for $i$ in the new distribution yielded by $f$.
Specifically, set 
$$\epsilon\coloneqq\min_{j \in T_{\q,i},\, j' \not \in T_{\q,i}} (q_{j'}p_{i,j}-q_j p_{i,j'})
\le \min_{j \in T_{\q,i},\, j' \not \in T_{\q,i}} \frac{q_{j'}p_{i,j}-q_j p_{i,j'}}{p_{i,j}+p_{i,j'}}.$$
Note that $\epsilon>0$, as $q_{j'}/p_{i,j'}>q_j/p_{i,j}$, by definition of critical alternatives.

By uniform continuity of $f$, there exists $\delta>0$ such that $\|f(P)-f(P')\|_1<2\epsilon$ for all $P'$ with $\|P-P'\|_1 \le \delta$. 
Set 
$$\lambda\coloneqq\min\left(1,\frac{\delta}{\|\p_i-\p^*_i\|_1}\right),$$ 
and define $\comb{P}$ as a profile identical to $P$ 
except that $i$ changes her peak from $\p_i$ to  
$\comb{\p}_i \coloneqq \lambda \p^*_i + (1-\lambda)\p_i$.
Note that 
$\|P-\comb{P}\|_1 =\lambda \|\p_i-\p^*_i\|_1 \leq \delta$,
so $\|\q-\comb{\q}\|_1 < 2\epsilon$,
where $\q=f(P)$ and $\comb{\q}=f(\comb{P})$.

The choice of $\epsilon$ ensures that $T_{\comb{\q},i} \subseteq T_{\q,i}$, as for arbitrary $j \in T_{\q,i}$ and $j' \not \in T_{\q,i}$ it holds that $\comb{q}_{j}<q_{j}+\epsilon$ and $\comb{q}_{j'}>q_{j'}-\epsilon$, so
\begin{align*}
    \frac{\comb{q}_{j'}}{p_{i,j'}}&>\frac{q_{j'}-\epsilon}{p_{i,j'}}
    \geq \frac{q_{j'}}{p_{i,j'}} - \frac{q_{j'}p_{i,j}-q_j p_{i,j'}}{p_{i,j'}(p_{i,j}+p_{i,j'})}
    =\frac{q_{j'}p_{i,j'}+q_j p_{i,j'}}{p_{i,j'}(p_{i,j}+p_{i,j'})}
    =\frac{q_{j'}+q_j}{p_{i,j}+p_{i,j'}}
    \\
    &=\frac{q_j p_{i,j}+q_{j'}p_{i,j}}{p_{i,j}(p_{i,j}+p_{i,j'})}
    =\frac{q_j}{p_{i,j}}+\frac{q_{j'}p_{i,j}-q_j p_{i,j'}}{p_{i,j}(p_{i,j}+p_{i,j'})}\geq\frac{q_j+\epsilon}{p_{i,j}}>\frac{\comb{q}_j}{p_{i,j}}.
\end{align*}
So every $j'$ which is not critical for $i$ under $\q$ cannot be critical for $i$ under $\comb{\q}$.
Therefore, 
\begin{align*}
    T_{\comb{\q},i} \subseteq T_{\q,i} = \comb{T}_{\q,i}.
\end{align*}

\emph{Proving that the outcome does not change.}
Consider a manipulation of agent $i$ who manipulates between reporting $\p_i$ and $\comb{\p}_i$.
Strategyproofness for $i$ implies 
both  $u_i(\q)\geq u_i(\comb{\q})$
and $\comb{u}_i(\comb{\q})\geq \comb{u}_i(\q)$.

The latter condition implies that, for every alternative $j \in T_{\q,i}$,
\begin{align*}
\frac{q_j}{\comb{p}_{i,j}}  &= \comb{u}_i(\q)  && \text{since $j\in T_{\q,i} = \comb{T}_{\q,i}$,}
\\
& \leq \comb{u}_i(\comb{\q})  && \text{by strategyproofness,}
\\
& \leq \frac{\comb{q}_j}{\comb{p}_{i,j}} && \text{by the definition of Leontief utilities.}
\end{align*}
So $q_j \leq \comb{q}_j$ for each alternative $j \in T_{\q,i}$. 
Together with $T_{\comb{\q},i} \subseteq T_{\q,i}$, this 
 implies $u_i(\q)\leq u_i(\comb{\q})$.
Therefore, $u_i(\q)=u_i(\comb{\q})$. 
Furthermore, if $\comb{u}_i(\comb{\q}) > \comb{u}_i(\q)$, then $\comb{q}_j>q_j$ for all $j \in \comb{T}_{\q,i} \supseteq T_{\comb{\q},i}$, which means that $u_i(\q)< u_i(\comb{\q})$, contradicting $u_i(\q)= u_i(\comb{\q})$. Thus, $\comb{u}_i(\comb{\q}) = \comb{u}_i(\q)$.  

Moreover, if the utility of some other agent $i'$ increases, group-strategyproofness is violated for the pair $\{i,i'\}$,
as this pair could profitably manipulate from $\q$ to $\comb{\q}$.
Similarly, if the utility of some other agent $i'$ decreases, group-strategyproofness is again violated for the pair $\{i,i'\}$,
as this pair could profitably manipulate from $\comb{\q}$ to $\q$.
Thus, $u_r(\q)=u_r(\comb{\q})$ for all $r \in N$.
Since $\q$ is efficient with respect to $P$, so is $\comb{\q}$.
By \Cref{lem:efficientunique}, $\q=\comb{\q}$.

Applying this argument repeatedly, we get a sequence of profiles $(P^k)$ with $P^0=P$ where $\p^k_i$ lies on the line $\lambda \p_i^*+(1-\lambda)\p_i$ for every $k$. It remains to show that $(\p^k)$ reaches $\p^*_i$ after a finite number of steps. For that, consider the expression in the definition of $\epsilon$:
\begin{align*}
\min_{j \in T_{\q,i},\, j' \not \in T_{\q,i}} (q_{j'}p_{i,j}-q_j p_{i,j'}).
\end{align*}
As $\p_i$ comes closer to $\p^*_i$, $p_{i,j}$ increases and $p_{i,j'}$ decreases while $\q$ and $T_{\q,i}$ stay the same, so overall the expression increases. Thus, we can take the $\epsilon$ (and the corresponding $\delta$) from the first step for every step. Furthermore, $\|P^k-P^{k+1}\|_1=\delta$ (unless $\lambda=1$, but then we have reached $\p^*_i$) implying that we reach $\p^*_i$ after at most $\lceil \|\p_i-\p^*_i\|_1/\delta \rceil$ steps; as we move on a line of length $\|P^k-P^{k'}\|_1=\sum_{\ell=k}^{k'-1}\|P^\ell-P^{\ell+1}\|_1$ for $k' \ge k$.

After the first agent has reached her desired peak $\p_i^*$, we turn to the next agent and repeat the procedure.
In that way, we eventually arrive at $P^*$.

\medskip
(b)
To see that $T^*_{\q,i}=T_{\q,i}$ for all $i\in N$, note that 
for every non-critical alternative $j \not\in T_{\q,i}$
we have $p^*_{i,j}=0$, so $j \not\in T^*_{\q,i}$.
Furthermore, for any critical alternative $j \in T_{\q,i}$
and any other $j' \in T_{\q,i}$,
\begin{align*}
    \frac{q_j}{p^*_{i,j}}=\frac{q_j \cdot \p_i(T_{\q,i})}{p_{i,j}}=\frac{q_{j'} \cdot \p_i(T_{\q,i})}{p_{i,j'}}=\frac{q_{j'}}{p^*_{i,j'}},
\end{align*}
so $j \in T^*_{\q,i}$. Therefore, $T_{\q,i}=T^*_{\q,i}$.
\end{proof}

\begin{lemma}\label{lem:keyprofs}
    Let $P^*$ be a profile and $\q$ be a distribution in which every agent values every non-critical alternative at $0$ ($j \not \in T^*_{\q,i}$ implies $p^*_{i,j}=0$ for any agent $i$). If $\q$ satisfies core fair share, then $\q=\nash(P^*)$. 
\end{lemma}

\begin{proof}
    Let $P^*$ be an arbitrary profile and let $\q \neq \nash(P^*)$ be a distribution such that $j \not \in T^*_{\q,i}$ implies $p^*_{i,j}=0$. In particular, $\q$ does not maximize Nash welfare. 
    By Lemma~11 of \citet{BGSS23a}, 
    there exists a group $N^-$ of agents 
    such that the total amount given to alternatives critical for some agent from $N^-$ is less than $|N^-|/n$.
    That is,

    \begin{align}\label{ineq:noequilibrium}
        \q(T^*_{\q, N^-}) < \frac{|N^-|}{n},
    \end{align}
where $T^*_{\q,N^-}\coloneqq\bigcup_{i \in N^-}T^*_{\q,i}$.

We will now show that core fair share is violated for this $N^-$.
This is clear if $\q(T^*_{\q, N^-}) = 0$, so assume that $\q(T^*_{\q, N^-}) > 0$.

Define a new distribution in which only alternatives in $T^*_{\q,N^-}$ are funded:
\begin{align*}
\q'&\coloneqq
\begin{cases}
    q_j / \q(T^*_{\q, N^-}) &\text{for $j \in T^*_{\q, N^-}$};\\
    0 &\text{for $j \not \in T^*_{\q, N^-}$}.
\end{cases}
\end{align*}
For every $i \in N^-$, as $p^*_{i,j}=0$ for $j \not \in T^*_{\q, N^-} \supseteq T^*_{\q, i}$,
the utility $u_i^*(\q')$ equals $u_i^*(\q)/\q(T^*_{\q,N^-})$, which is larger than $u_i^*(\q)/ (|N^-|/n)$ by \eqref{ineq:noequilibrium}.
Therefore, 
the utility 
$u^*_i\left((|N^-|/n)\q'+\left(1-|N^-|/n\right)\q''\right)$ is at least $(|N^-|/n)u^*_i(\q') > u_i^*(\q)$ for every $\q'' \in \Delta^m$,
contradicting core fair share for $N^-$.
\end{proof}

\begin{lemma}\label{lem:keyprofseq}
Let $P^*$ and $P$ be profiles where $T^*_{\q,i}=T_{\q,i}$ for $\q=\nash[P^*]$ and all $i \in N$. Then, $\nash[P]=\q$.
\end{lemma}

\begin{proof}
    As $\q$ maximizes Nash welfare in $P^*$, by \Cref{lem:eq-iff-critical} there exists a decomposition $(\s_i)_{i \in N}$ such that $s_{i,j}=0$ for every $i \in N$ and $j \not \in T^*_{\q,i}$. Due to $T^*_{\q,i}=T_{\q,i}$, the same decomposition proves that $\q$ also maximizes Nash welfare in $P$ by \Cref{lem:eq-iff-critical}, thus $\nash[P]=\q$.
\end{proof}

\begin{proof}[Proof of \Cref{thm:Nashchar}]
    Let $P$ be an arbitrary profile, and  
    $P^*$ a modified profile defined as in \Cref{lem:keyprofsdistr}. Then, 
    \begin{align*}
        f(P)\overset{\Cref{lem:keyprofsdistr}}{=}f(P^*)\overset{\Cref{lem:keyprofs}}{=}\nash[P^*]\overset{\Cref{lem:keyprofseq}}{=}\nash[P],
    \end{align*}
    where \Cref{lem:keyprofseq} uses the fact that the sets of critical alternatives under $\q$ did not change when moving from $P$ to $P^*$.
\end{proof}

\section{Proofs for \Cref{sec:m=2}} \label{sec:proofsm=2}

We start with a characterization for a single agent; this will be used in further characterizations.

\begin{lemma}
\label{lem:m2n1}
For $m=2$ and $n=1$, a mechanism $f$ on 
a domain $\mathcal{U} \subseteq \mathcal{U}^{SP}$
is continuous and strategyproof if and only if there exist $\alpha_0 \le \alpha_1$ in $[0,1]$ such that
    \begin{align*}
        f(p)=\med(p, \alpha_0, \alpha_{1}).
    \end{align*}
\end{lemma}
\begin{proof}
The ``if'' direction is obvious; we focus on the ``only if''.

Let $f$ be a continuous strategyproof mechanism for $n=1$ and any $m$.
Let $S \coloneqq f(\Delta^m)$ be the image of $f$.
By continuity, $S$ is a closed set.
By strategyproofness, for all $\p\in \Delta^m$ we have $\displaystyle f(\p)\in \arg\max_{\q\in S}u(\q)$.
Moreover, $S$ must be convex, since if a segment with endpoints in $S$ is not contained in $S$, then some internal point of this segment would be a discontinuity point of $\arg\max$.

For $n=1$ and $m=2$, this boils down to $S$ being a closed interval, $S = [\alpha_0,\alpha_1]$ for some $\alpha_0 \le \alpha_1$ in $[0,1]$,
and $f$ being a function that maps each $p$ to the point nearest to $p$ in $[\alpha_0,\alpha_1]$.
This is equivalent to $f(p) = \med(p, \alpha_0,\alpha_1)$.
\end{proof}

Note that continuity is essential for the above characterization. For example, the following discontinuous mechanism is strategyproof:
\begin{align*}
    f(p) \coloneqq \begin{cases}
	    0  & p<0.5; 
	    \\
	    1  & p\geq 0.5.
    \end{cases}
\end{align*}

We now move on to mechanisms for any number of agents. The following lemma shows that, for $m=2$, any continuous strategyproof mechanism is completely determined by its outcomes on single-minded profiles.

\begin{lemma}
\label{lem:m2} 
For $m=2$ and arbitrary domain  $\mathcal{U} \subseteq \mathcal{U}^{SP}$, 
if two continuous and strategyproof mechanisms yield the same distribution for all single-minded profiles, then they yield the same distribution for all profiles.
\end{lemma}
\begin{proof}
Let $f$ and $g$ be two continuous strategyproof mechanisms.
Let $P$ a profile for which $f(P)\neq g(P)$. 
We prove that there is a single-minded profile $P'$ for which $f(P')\neq g(P')$.

At a high level, the proof works as follows. Step by step, each agent with peak on the left side of $f(P)$ moves her peak closer and closer to $0$ and each agent with peak on the right side moves to~$1$. Continuity and strategyproofness imply that $f(P)$ cannot change in the process. Finally, for all agents with peaks at $f(P)$, move their peaks to the alternative that is not ``separated'' from the peak by $g(P)$. In the process, $f(P)$ can only move further away from $g(P)$.

In detail, assume that $f(P)<g(P)$; the case $f(P)>g(P)$ can be handled analogously.
Denote $q \coloneqq f(P)$.

Partition the set of agents into four groups: $N = N^{01} \cup N^- \cup N^= \cup N^+ $, where
$N^{01}=\{i \in N:p_i \in \{0,1\}\}$, $N^-=\{i \in N \setminus N^{01}: p_i<q\}$, $N^==\{i \in N \setminus N^{01}: p_i=q\}$, and $N^+=\{i \in N \setminus N^{01}: p_i>q\}$. Our overall goal is to ``move'' all agents to $N^{01}$ while keeping the chosen distribution different from $g(P)$.

Take any agent $i \in N^-$, and consider the function $F:[0,1] \to [0,1]$ defined by $F(p)\coloneqq f(p,P_{-i})$. 
Since $f$ is continuous and strategyproof, so is $F$, as a mechanism for a single agent. Hence, by \Cref{lem:m2n1}, 
$F(p)=\med(p, \alpha_0, \alpha_1)$ for some constants $\alpha_0\leq \alpha_1$.
Note that $F(p_i)=f(P) > p_i$ as $i\in N^-$, so $p_i<\med(p_i,\alpha_0,\alpha_1)$. 
The median properties imply that $F(p)=F(p_i)$ also for all $p < p_i$. 
In particular, $F(0)=F(p_i)= f(P)$.

Denote the profile where agent $i$ changed her peak to $0$ by $P^{\{i\}}$; then $f(P^{\{i\}}) = F(0) =f(P)$. The same argument applies to all other agents from $N^-$, so $f(P^{N^-})=f(P)$, where $P^{N^-}$ denotes the profile resulting from $P$ after all agents in $N^-$ moved their peak to $0$. 
Also, $g(P^{N^-})=g(P)$, as all agents from $N^-$ were also on the left side of $g(P)$ due to $f(P)<g(P)$, so moving them further left does not change the distribution returned by $g$.
    
For an agent $i \in N^+$, define $F(p) \coloneqq f(p, P^{N^-}_{-i})$. One can show analogously that $F(p)=F(P_i)=f(P^{N^-})= f(P)$ for all $p \ge p_i$, so the outcome remains $f(P)$ when $i$ moves her peak to $1$. Therefore, $f(P^{N^- \cup N^+})=f(P)$, where $P^{N^- \cup N^+}$ denotes the profile resulting from $P$ after all agents in $N^-$ moved their peak to $0$ and all agents in $N^+$ moved their peak to $1$.
Also, $g(P^{N^- \cup N^+}) \ge g(P^{N^-})$ as moving peaks to the right can only increase the median returned by $g$.
Therefore, $f(P^{N^- \cup N^+}) < g(P^{N^- \cup N^+})$ still holds.

We now consider an agent $i \in N^=$, for whom $p_i=f(P)<g(P) \le g(P^{N^- \cup N^+})$. 
Define $F'(p)\coloneqq f(p,P^{N^- \cup N^+}_{-i})$. 
As it is continuous and strategyproof, 
\Cref{lem:m2n1} implies that
$F'(p) = \med(p, \alpha_0, \alpha_1)$ for some constants $\alpha_0\leq \alpha_1$.
As the median is a weakly monotone function of its arguments, $F'(p) \le F'(p_i)$ for $p \le p_i$.
Thus, with $P^{N^- \cup N^+ \cup \{i\}}$ denoting the profile where agent $i$ moved her peak to $0$, $f(P^{N^- \cup N^+ \cup \{i\}}) \le f(P^{N^- \cup N^+})<g(P^{N^- \cup N^+})=g(P^{N^- \cup N^+ \cup \{i\}})$. If $f(P^{N^- \cup N^+ \cup \{i\}}) = f(P^{N^- \cup N^+})$, repeat the procedure with the next agent from $N^=$. If $f(P^{N^- \cup N^+ \cup \{i\}}) < f(P^{N^- \cup N^+})$, all remaining agents from $N^=$ now have their peak on the right side of $f(P^{N^- \cup N^+ \cup \{i\}})$ and can move their peak to~$1$ without changing the chosen distribution, as in the case $i\in N^+$. 
Again, the outcome from $g$ can only move to the right or stay fixed. 

Let $P^{N^- \cup N^+ \cup N^=}$ denote the profile after all agents in $N^- \cup N^+ \cup N^=$ have moved their peaks. This profile is single-minded, as all agents have their peaks at $0$ or $1$, and $f(P^{N^- \cup N^+ \cup N^=})<g(P^{N^- \cup N^+ \cup N^=})$, as required. 
\end{proof}

\begin{remark}
    \citet{BoJo83a} considered a property called \emph{uncompromisingness}, which states that the outcome cannot change when agents from $N^-$ and $N^+$ move their peaks to $0$ and~$1$, respectively (i.e., there is no compromise with agents who express extreme preferences). They showed that uncompromisingness implies continuity. By contrast, we assume continuity and obtain uncompromisingness.
    As a result, all our characterizations hold if we replace continuity with uncompromisingness.
\end{remark}

Using \Cref{lem:m2}, we can now prove several characterizations.

\subsection{Characterizing generalized median rules}

\Moulsubdomain*

\begin{proof}
The ``if'' direction is obvious; we focus on the ``only if''.

For any $k\in\{0,\ldots,n\}$, let $P_k$ be a single-minded profile in which some $k$ agents have their peak at $1$ and the other $n-k$ have their peak
at $0$. Let $\alpha_k \coloneqq f(P_k)$; due to the anonymity of $f$, it holds that $\alpha_k$ does not depend on the selection of $P_k$.

Since $f$ is strategyproof, $\alpha_k \leq \alpha_{k+1}$ for all $k\in\{0,\ldots,n-1\}$; otherwise, in profile $P_k$, some agent with peak at $0$ could gain from reporting a peak at $1$.

Let $g(P) \coloneqq \med(p_1,\dots,p_n,\alpha_0, \dots, \alpha_{n})$. Then for any $k\in\{0,\ldots,n\}$, $g(P_k) = \alpha_k$, as $n$ arguments of the median are at most $\alpha_k$ and $n$ arguments are at least $\alpha_k$.
This means that $f$ and $g$ agree on all single-minded profiles. By \Cref{lem:m2}, $f\equiv g$.
\end{proof}

Two additional characterizations can be obtained in a similar way (in the first, we add efficiency, while in the second, we remove anonymity):
\begin{enumerate}
    \item A continuous mechanism $f$ satisfies anonymity,  strategyproofness, and efficiency if and only if there exist $\alpha_1 \le \dots \le \alpha_{n-1}$ in $[0,1]$ such that
    \begin{align*}
        f(P)=\med(p_1,\dots,p_n,\alpha_1, \dots, \alpha_{n-1}).
    \end{align*}
    \item A continuous mechanism $f$ satisfies strategyproofness if and only if there exist 
 $2^n$ constants, $\alpha_G \in [0,1]$ for all $G\subseteq N$, such that
    \begin{align*}
        f(P) = 
        \max_{G\subseteq N}
        \min(\alpha_G, \min_{i\in G} p_i).
        \end{align*}
\end{enumerate}

\subsection{Characterizing the uniform phantom rule}

\unifphantomchar*

\begin{proof}
Let $g$ be the uniform phantom mechanism.
Then $g$ is proportional.
Since proportionality completely specifies the outcomes for all single-minded profiles, any proportional mechanism agrees with $g$ on all single-minded profiles.
By \Cref{lem:m2}, any such mechanism must equal $g$.
\end{proof}

For $m=2$, an outcome is efficient for a profile $P$ if and only if it is range-respecting, which means that the uniform phantom rule is efficient. Thus, for only two alternatives, there is a unique way to aggregate utilities in an efficient, strategyproof, and fair manner, even without knowledge of the specific underlying utility model. 

\section{Leximin-Leontief preferences}
\label{sec:leximin}

Leontief utilities, as defined in \Cref{sub:mq-utils}, assume that agents rank distributions only by the smallest ratio, $\min_{j\in M_i}q_j/p_{i,j}$. 
In this section, we assume that agents rank distributions with the same smallest ratio by the second-smallest ratio, and 
distributions with the same smallest and second-smallest ratio by the third-smallest ratio, and so on. We call these preferences \emph{Leximin-Leontief preferences}. We denote the strict Leximin-Leontief preferences of each agent $i$ by $\succ^{lex}_i$, and the weak relation by $\succeq^{lex}_i$.
When we want to emphasize that the leximin relation uses a specific peak $\p_i$, we write 
$\succ^{lex}_{\p_i}$ and $\succeq^{lex}_{\p_i}$.

We still define the \nash rule based on the minimum ratio only, which we continue to denote by $u_i(\q)$. Therefore, the \nash distribution remains a continuous function of the peaks (even though the Leximin-Leontief preferences are not continuous).
However, the change of preferences may potentially affect some properties of the rule.
In particular, $\q \succeq^{lex}_i \q'$  implies $u_i(\q)\geq u_i(\q')$, but for the strict relation the opposite direction is true:
$u_i(\q)> u_i(\q')$ implies $\q \succ^{lex}_i \q'$.
Therefore, properties defined by the weak relation only, such as strategyproofness, are  stronger with Leximin-Leontief preferences than with Leontief utilities. 
However, properties defined by both the weak and the strict relations, such as group-strategyproofness, core fair share, and efficiency, are not a priori stronger or weaker with Leximin-Leontief preferences than with Leontief utilities.

First, we claim that \Cref{lem:efficient-iff-critical} still holds, where the critical alternatives are defined as in \Cref{def:critical} (based on the minimum ratio only).
\begin{lemma}
\label{lem:efficient-iff-critical-leximin}
With Leximin-Leontief preferences, 
an outcome $\q$ is efficient if and only if every alternative  $j$ with $q_j>0$ is critical for some agent. 
\end{lemma}

\begin{proof}[Proof sketch]
$\Rightarrow$: Suppose that some alternative $j$ with $q_j>0$ is not critical for any agent. 
We can construct a new outcome $\q'$ by removing a sufficiently small amount from $j$ and distributing it equally among all other alternatives.
This increases $\min_j q_j/p_{i,j}$ for all agents, and thus makes the new distribution strictly leximin-better for all agents.
Hence, $\q$ is not efficient.

$\Leftarrow$: Suppose that every alternative $j$ with $q_j>0$ is critical for some agent.
Let $\q'$ be any outcome different than $\q$, and let $y$ be an alternative with $q'_y<q_y$.
As $q_y>0$, the assumption implies that $y$ is critical to some agent; denote one such agent by $i_y$. 
Then
\begin{align*}
 \min_j \frac{q'_j}{p_{i_y,j}} 
 \leq 
\frac{q'_y}{p_{i_y,y}} 
 &< 
\frac{q_y}{p_{i_y,y}} 
\\
&= 
\min_j \frac{q_j}{p_{i_y,j}}\text,  && \text{(as $y$ is critical for $i_y$ under $\q$)}
\end{align*}
so $\q'$ is leximin-worse for $i$ than $\q$. Hence, $\q'$ does not Pareto-dominate $\q$.
This holds for all $\q'$, which implies that $\q$ is efficient.
\end{proof}
\Cref{lem:efficient-iff-critical-leximin} implies that $\q$ is efficient for Leximin-Leontief preferences if and only if it is efficient for the corresponding Leontief utilities. In particular, \nash remains efficient.
Moreover, \nash is still neutral and \Cref{lem:one-sided-range-respect} (efficiency implies one-sided range-respect) remains valid as well.

Next, we show that \nash remains group-strategyproof too. We need a lemma.

\begin{lemma}
\label{lem:twodistributions}
Let $\q'$ and $\q''$ be two distributions, and $i\in N$ an agent.
If $\q'' \succeq^{lex}_i \q'$, 
then every alternative in $T_{\q',i}$ receives at least as much in $\q''$ as in $\q'$, that is,
$q''_y \geq q'_y$ for all $y\in T_{\q',i}$.
\end{lemma}

\begin{proof}
For every alternative $y \in T_{\q',i}$:
 \begin{align*}
 q'_y &= p_{i,y}\cdot u_i(\q') && \text{(as $y$ is critical for $i$ under $\q'$)}
 \\
 &\leq p_{i,y}\cdot u_i(\q'') && \text{(since $\q'' \succeq_i^{lex} \q'$ implies $u_i(\q'')\geq u_i(\q')$)}
 \\
 &=p_{i,y}\cdot \min_{j \in M_i}\frac{q''_j}{p_{i,j}} 
 && \text{(by definition of $u_i$)}
\\
& \leq p_{i,y}\cdot \frac{q''_y}{p_{i,y}} && \text{(since $y \in T_{\q',i}\subseteq M_i$)}
 \\
 & = q''_y,
 \end{align*}
 completing the proof.
\end{proof}

\begin{theorem}\label{thm:nashgspleximin}
With Leximin-Leontief preferences, \nash is group-strategyproof.
\end{theorem}

\begin{proof}
Assume for contradiction that there exist profiles $P$ and $P'$ with \nash distributions $\q\neq \q'$ respectively, and an inclusion-maximal group of agents $G\subseteq N$ which do not lose from the manipulation from $P$ to $P'$.
Let $T_{\q,G}\coloneqq\bigcup_{i \in G}T_{\q,i}$ be the set of alternatives critical to at least one agent from $G$ under $\q$.
As no agent from $G$ loses from the manipulation, 
\Cref{lem:twodistributions} implies that 
$q'_x \ge q_x$ for all $x \in T_{\q,G}$.

As $\q'\neq \q$, there is an alternative $y\in M$ for which $q'_y > q_y$. Denote $B\coloneqq T_{\q,G} \cup \{y\}$
(it is possible that $y\in T_{\q,G}$).
Then, $\q'(B)>\q(B)$.

We now consider the decompositions of $\q$ and $\q'$
guaranteed to exist by \Cref{lem:eq-iff-critical}. 
Since $\q'(B)>\q(B)$, there exists an agent $j\in N$ who contributes more to $B$ in the decomposition of $\q'$ than in the decomposition of $\q$. 
This implies that, in the decomposition of $\q$, agent $j$ contributes some of her share of $1/n$ to alternatives not in $B$. 
It follows that 
$T_{\q,j} \not \subseteq B$, so $j \not\in G$,
and thus $u_j=u'_j$ (as $j$ is not a part of the manipulating group). 

In the decomposition of $\q'$, agent $j$ must contribute a positive amount to some alternative $x\in B$, which means that $x$ is critical for $j$ under $\q'$. Since $u_j=u'_j$, we have $u_j(\q')=u'_j(\q')=q'_x/p'_{j,x}\geq q_x/p'_{j,x} \geq u_j(\q)$. Therefore, all agents in $G \cup \{j\}$ do not lose from the manipulation, which contradicts the maximality of $G$.
\end{proof}

We now extend \Cref{prop:nash-cfs} to Leximin-Leontief preferences.

\begin{proposition}
\label{prop:nash-cfs-leximin}
With Leximin-Leontief preferences,
    \nash satisfies core fair share.
\end{proposition}

\begin{proof}
Assume for contradiction that there exists $P \in \mathcal{P}$  such that $\q\coloneqq\nash[P]$ does not satisfy core fair share 
for some $G \subseteq N$. Then, there exists $\q' \in \Delta^m$ such that, for every $\q'' \in \Delta^m$,
\begin{align*}
(|G|/n)\q'+(1-|G|/n)\q''  \succ^{lex}_i \q && \text{for all $i \in G$.}
\end{align*}

Let $T_{\q,G}\coloneqq\bigcup_{i \in G}T_{\q,i}$ be the set of alternatives critical to at least one agent from $G$.

Note that $T_{\q,G}=M$ cannot hold. Otherwise, by Lemma \ref{lem:efficient-iff-critical-leximin}, $\q$ would be efficient not only for~$N$ but already for $G$, contradicting that $\q$ does not satisfy core fair share for $G$.
Therefore, there exists 
a distribution $\q''$ with $q''_j=0$
for every $j \in T_{\q,G}$.
Choosing such a distribution $\q''$ shows that $(|G|/n)q'_j \ge q_j$;
otherwise some agent from $G$ for whom $j$ is critical would have a smaller utility.
Thus, $\q(T_{\q,G})\coloneqq\sum_{j \in T_{\q,G}}q_j 
\le (|G|/n)\cdot \sum_{j \in T_{\q,G}}q'_j
\le |G|/n$. 

By \Cref{lem:eq-iff-critical}, 
the \nash distribution can be decomposed in such a way that
every agent from $G$ only contributes her share of $1/n$ 
to alternatives in $T_{\q,G}$. Thus, $\q(T_{\q,G}) \ge |G|/n$. All in all, $\q(T_{\q,G}) = |G|/n$
and $(|G|/n)q'_j=q_j$ for every $j \in T_{\q,G}$. 
But this also implies that 
$(|G|/n)\cdot \q'(T_{\q,G}) = |G|/n$, so $\q'(T_{\q,G}) = 1$.
This means that $\q'$ only allocates to alternatives in $T_{\q,G}$. 
As the allocation to alternatives in $T_{\q,G}$  is the same in $\q$ and $(|G|/n)\q'$, no agent in $G$ can have a better leximin vector in $(|G|/n)\q'$ than in $\q$.
\end{proof}

We now consider the characterization (\Cref{thm:Nashchar}) for Leximin-Leontief preferences. It turns out that group-strategyproofness can be weakened to strategyproofness.

As in the proof of \Cref{thm:Nashchar}, to show the statement, we would like to change $P$ gradually to $P^*$, where each agent's peak puts $0$ on non-critical alternatives. However, in order to exploit the fact that Leximin-Leontief preferences constitute a refinement of Leontief utilities, which allows us to weaken group-strategyproofness to strategyproofness, we need to adapt the proof. 
We first show that $f$ coincides with \nash on all \emph{strictly-positive profiles}, that is, all profiles 
$P \in \mathcal{P}^+$, where $\mathcal{P}^+\coloneqq\{P \in \mathcal{P}: p_{i,j}>0 \text{ for all }i \in N,j \in M\}$.

\begin{lemma}\label{lem:keyprofsdistrlex}
If $f$ is a continuous mechanism satisfying strategyproofness, then when moving from $P$ to $P^*$,

(a) the outcome does not change, that is, $f(P^*)=f(P)=\q$ for $P \in \mathcal{P}^+$;

(b) the sets of critical alternatives do not change, that is, $T_{\q,i}=T^*_{\q,i}$ for every $i \in N$.
\end{lemma}

\begin{proof}
We first show the statement for a slightly perturbed $\tilde{P}^*_{\tilde{\epsilon}}$ with $\tilde{\p}^*_i=(1-\tilde{\epsilon})\p^*_i+\tilde{\epsilon}\p_i$ and arbitrary small but fixed $\tilde{\epsilon}>0$. 
Note that $P^*$ may not be in $\mathcal{P}^+$,
but $\tilde{P}^*_{\tilde{\epsilon}}$ is always in $\mathcal{P}^+$.
Note also that, for an agent $i$ with $T_{\q,i}=M$, it holds that $\p_i=\tilde{\p}^*_i = \p_i^*$.

Again, we move the peak of each agent in turn.
For each agent $i$, we change $\p_i$ towards~$\tilde{\p}^*_i$ gradually, to some  $\comb{\p}_i \coloneqq \lambda \tilde{\p}_i^*+(1-\lambda)\p_i$, for some $\lambda \in [0,1]$ to be computed later. 
Then we proceed along this line until we reach $\lambda=1$ and $\tilde{\p}^*_i$.

Given the outcome $\q = f(P)$, we partition the alternatives of each agent $i$ into \emph{critical classes}, i.e., subsets with the same ratio $q_j/p_{i,j}$. 
Here we use the fact that $P \in\mathcal{P}^+$, so $p_{i,j}>0$ for all $i,j$.
Denote the subset with the smallest ratio by $T_{\q,i,1} \equiv T_{\q,i}$, the subset with the second-smallest ratio by $T_{\q,i,2}$, and so on, up to $T_{\q,i,w}$, where $w$ is the number of different ratios.
Also, for $r\in [w]$, denote $T_{\q,i,\leq r} \coloneqq T_{\q, i,1} \cup \cdots \cup T_{\q, i,r}$, and define $T_{\q,i,> r}$ analogously.

As $\comb{\p}_i$ lies along the line between $\p_i$ and $\p^*_i$, the change from $\p_i$ to $\comb{\p}_i$ has a simple structure:
\begin{itemize}
\item 
$\comb{p}_{i,j} > p_{i,j}$ for all $j\in T_{\q,i,1}$,
and the ratio $\comb{p}_{i,j} / p_{i,j} = \lambda (1-\tilde{\epsilon})/\p_i(T_{\q, i,1})+\lambda \tilde{\epsilon} + (1-\lambda) =: \lambda^+$, a constant independent of $j$;
\item 
$\comb{p}_{i,j} < p_{i,j}$ for all $j  \in T_{\q, i,> 1}$, 
and the ratio 
$\comb{p}_{i,j} / p_{i,j} = \lambda \tilde{\epsilon}+(1-\lambda) =: \lambda^-$, again independent of~$j$.
\end{itemize}

\emph{Computing $\lambda$.}
We pick $\lambda$ sufficiently small such that no new alternative becomes critical for $i$, and moreover, critical classes do not mix, i.e., $q_j'/p_{i,j'}>q_{j}/p_{i,j}$ implies $\comb{q}_{j'}/p_{i,j'}>\comb{q}_{j}/p_{i,j}$ for all $j,j' \in M$.
Specifically, set 
$$\epsilon\coloneqq\min_{j \in T_{\q,i,r},\, j' \in T_{\q,i,s}} \frac{q_{j'}p_{i,j}-q_j p_{i,j'}}{p_{i,j}+p_{i,j'}}$$ where the minimum is taken over all $r,s \in [w]$ and $s>r$.
Note that $\epsilon>0$, as $q_{j'}/p_{i,j'}>q_j/p_{i,j}$, by definition of critical classes.

By uniform continuity of $f$, there exists $\delta>0$ such that $\|f(P)-f(P')\|_1<2\epsilon$ for all $P'$ with $\|P-P'\|_1 \le \delta$. 
Set 
$$\lambda\coloneqq\min\left(1,\frac{\delta}{\|\p_i-\tilde{\p}^*_i\|_1}\right),$$ 
and define $\comb{P}$ as a profile identical to $P$ 
except that $i$ changes her peak from $\p_i$ to  
$\comb{\p}_i \coloneqq \lambda \tilde{\p}^*_i + (1-\lambda)\p_i$.
Note that 
$\|P-\comb{P}\|_1 =\lambda \|\p_i-\tilde{\p}^*_i\|_1 \leq \delta$,
so $\|\q-\comb{\q}\|_1 < 2\epsilon$,
where $\q=f(P)$ and $\comb{\q}=f(\comb{P})$.

The choice of $\epsilon$ ensures that for arbitrary $r,s \in [w]$ with $s>r$, $j \in T_{\q,i,r}$, and $j' \in T_{\q,i,s}$,
\begin{align*}
    \frac{\comb{q}_{j'}}{p_{i,j'}}&>\frac{q_{j'}-\epsilon}{p_{i,j'}}
    \geq \frac{q_{j'}}{p_{i,j'}} - \frac{q_{j'}p_{i,j}-q_j p_{i,j'}}{p_{i,j'}(p_{i,j}+p_{i,j'})}
    =\frac{q_{j'}p_{i,j'}+q_j p_{i,j'}}{p_{i,j'}(p_{i,j}+p_{i,j'})}
    =\frac{q_{j'}+q_j}{p_{i,j}+p_{i,j'}}
    \\
    &=\frac{q_j p_{i,j}+q_{j'}p_{i,j}}{p_{i,j}(p_{i,j}+p_{i,j'})}
    =\frac{q_j}{p_{i,j}}+\frac{q_{j'}p_{i,j}-q_j p_{i,j'}}{p_{i,j}(p_{i,j}+p_{i,j'})}\geq\frac{q_j+\epsilon}{p_{i,j}}>\frac{\comb{q}_j}{p_{i,j}}.
\end{align*}
Therefore, we have $T_{\q,i,r} = \comb{T}_{\q,i,r}$ for all $r \in [w]$.

\emph{Proving that the outcome does not change.}
Consider a manipulation of agent $i$ who manipulates between reporting $\p_i$ and $\comb{\p}_i$.
Strategyproofness for agent $i$ implies 
both  $\q \succeq^{lex}_{\p_i} \comb{\q}$ 
and $\comb{\q} \succeq^{lex}_{\comb{\p}_i} \q$.

We now prove, by induction on $r$, that $q_j=\comb{q}_j$ for all $j\in T_{\q,i,r}$.
For the base case $r=1$, consider the alternatives in $T_{\q,i,1}$.
\begin{itemize}
\item
As all alternatives in $T_{\q,i,1}$ are at the bottom of the ordering by ${q}_j/p_{i,j}$ (by definition) as well as by $\comb{q}_j/p_{i,j}$ (by the choice of $\epsilon$), the relation $\q \succeq^{lex}_{\p_i} \comb{\q}$  
implies the same relation among the sub-vectors corresponding to the alternatives in $T_{\q,i,1}$, that is, 
\begin{align}
\label{eq:lexpi}
    [q_j \mid j\in T_{\q,i,1}] \succeq^{lex}_{\p_i} [\comb{q}_j \mid j\in T_{\q,i,1}].
\end{align}
\item 
Similarly, all alternatives 
in $\comb{T}_{\q,i,1} = T_{\q,i,1}$ are at the bottom of the ordering by 
$q_j/\comb{p}_{i,j}$ by construction.
Therefore, the relation $\comb{\q} \succeq^{lex}_{\comb{\p}_i} \q$ 
implies 
\begin{align}
\label{eq:lexpihat}
    [\comb{q}_j \mid j\in T_{\q,i,1}] \succeq^{lex}_{\comb{\p}_i} [q_j \mid j\in T_{\q,i,1}].
\end{align}
\item 
But since $\comb{p}_{i,j}$ differs from $p_{i,j}$
by a constant factor $\lambda^+$ for all $j\in T_{\q,i,1}$, 
\eqref{eq:lexpihat} implies the same inequality with $\succeq^{lex}_{\p_i}$ instead of $\succeq^{lex}_{\comb{\p}_i}$.
Combining this with \eqref{eq:lexpi}, we get
\begin{align*}
[\comb{q}_j \mid j\in T_{\q,i,1}] \simeq^{lex}_{\p_i} [q_j \mid j\in T_{\q,i,1}].
\end{align*}
As $q_j/p_{i,j}$ is constant for $j \in T_{\q,i,1}$, lexicographic equivalence with respect to $\p_i$ implies $q_j/p_{i,j}=\comb{q}_j/p_{i,j}$ for all $j \in T_{\q,i,1}$.
Thus, $q_j = \comb{q}_j$ must hold for all 
$j\in T_{\q,i,1}$.
\end{itemize}

For the induction step, assume that $q_j = \comb{q}_j$ holds for all 
$j\in T_{\q,i,\le r}$, for some $r\in[w-1]$.
Next, consider the alternatives in $T_{\q,i,r+1}=\comb{T}_{\q,i,r+1}$.

\begin{itemize}
\item
As $q_j = \comb{q}_j$ holds for all other alternatives with smaller ratios, the relation $\q \succeq^{lex}_{\p_i} \comb{\q}$  
implies the same relation for the subset $T_{\q,i,r+1}$,
that is,
$[q_j \mid j\in T_{\q,i,r+1}] \succeq^{lex}_{\p_i} [\comb{q}_j \mid j\in T_{\q,i,r+1}]$.
\item 
Similarly, the relation $\comb{\q} \succeq^{lex}_{\comb{\p}_i} \q$ implies $[\comb{q}_j \mid j\in T_{\q,i,r+1}] \succeq^{lex}_{\comb{\p}_i} [q_j \mid j\in T_{\q,i,r+1}]$.
\item 
But since $\comb{p}_{i,j}$ differs from $p_{i,j}$
by a constant factor $\lambda^-$ for all $j\in T_{\q,i,r+1}$, 
$\comb{\q} \succeq^{lex}_{\comb{\p}_i} \q$ implies 
$\comb{\q} \succeq^{lex}_{\p_i} \q$,
so all in all 
$[\comb{q}_j \mid j\in T_{\q,i,r+1}] \simeq^{lex}_{\p_i} [q_j \mid j\in T_{\q,i,r+1}]$ must hold.
This means that $q_j = \comb{q}_j$ must hold for all 
$j\in T_{\q,i,r+1}$.
\end{itemize}

Therefore, $\comb{\q}=\q$.

Applying this argument repeatedly, we get a sequence of profiles $(P^k)$ with $P^0=P$ where $\p^k_i$ lies on the line $\lambda \tilde{\p}_i^*+(1-\lambda)\p_i$ for every $k$. It remains to show that $(\p^k)$ reaches $\tilde{\p}^*_i$ after a finite number of steps. For that, consider the expression in the definition of $\epsilon$:
\begin{align*}
\min_{j \in T_{\q,i,r},\, j' \in T_{\q,i,s}} \frac{q_{j'}p_{i,j}-q_j p_{i,j'}}{p_{i,j}+p_{i,j'}}.
\end{align*}
For $r=1$, as $\p_i$ comes closer to $\tilde{\p}^*_i$, $p_{i,j}$ increases and $p_{i,j'}$ decreases while $\q$ and the critical classes stay the same, so overall the expression increases. For $s>r>1$, note that $p_{i,j}$ and $p_{i,j'}$ both decrease by the same factor $\lambda^-$ while $\q$ and the critical classes stay the same. Thus, we can take the $\epsilon$ (and the corresponding $\delta$) from the first step for every step. Furthermore, $\|P^k-P^{k+1}\|_1=\delta$ (unless $\lambda=1$, but then we have reached $\tilde{\p}^*_i$) implying that we reach $\tilde{\p}^*_i$ after at most $\lceil \|\p_i-\tilde{\p}^*_i\|_1/\delta \rceil$ steps; as we move on a line of length $\|P^k-P^{k'}\|_1=\sum_{\ell=k}^{k'-1}\|P^\ell-P^{\ell+1}\|_1$ for $k' \ge k$.

After the first agent has reached her desired peak $\tilde{\p}_i^*$, we turn to the next agent and repeat the procedure.
In that way, we eventually arrive at $\tilde{P}^*_{\tilde{\epsilon}}$.

As $\tilde{\epsilon}$ was chosen arbitrarily and we have $\lim_{\tilde{\epsilon} \to 0} P^*_{\tilde{\epsilon}}=P^*$ for arbitrary $P \in \mathcal{P}^+$, continuity of $f$ implies $f(P)=f(P^*)$.

Statement (b) now follows analogously to the one in \Cref{lem:keyprofsdistr}.
\end{proof}

\begin{proof}[Proof of \Cref{thm:Nashcharlex}]
As \Cref{lem:keyprofs,lem:keyprofseq} still hold for Leximin-Leontief preferences, $f(P)=\nash[P]$ for all $P \in \mathcal{P}^+$.
Noting that $\mathcal{P}^+$ is dense in $\mathcal{P}$ (and $f$ and $\nash$ are continuous), $f$ has to coincide with $\nash$ on all profiles in $\mathcal{P}$.
\end{proof}
\end{document}